\documentclass[10pt]{sig-alternate-10pt}

\usepackage{epsf}
\usepackage{epsfig}
\usepackage{mathrsfs}
\usepackage{subfigure}
\usepackage{graphics}
\usepackage{graphicx}
\usepackage{epstopdf}

\newcommand{\comment}[1]{}
\renewcommand{\textfloatsep}{3pt}
\begin{document}

\title{
	A measurement driven, 802.11 anti-jamming system \\ 
}
\author{ 
Konstantinos Pelechrinis$^*$, 
Ioannis Broustis$^*$, 
Srikanth V. Krishnamurthy$^*$, 
Christos Gkantsidis$^{\dagger}$ \\
\affaddr{$^*${University of California,  Riverside ~~~~~~~~~~~~~~~~~~~~~~ $^{\dagger}$Microsoft Research, Cambridge, UK}} \\  
\normalfont{\{kpele, broustis, krish\}@cs.ucr.edu  ~~~~~~~~~~~~~~~~~~~~~~~~~~~~~~~~ christos.gkantsidis@microsoft.com}\\
} 


\date{}
\maketitle
 
\begin{abstract}

Dense, unmanaged 802.11 deployments tempt saboteurs 
into launching jamming attacks by injecting malicious interference. 
Nowadays, jammers can be portable devices that transmit intermittently at low power  in order to conserve energy. 
In this paper, we first conduct extensive experiments on an indoor 802.11 network to 
assess the ability of two physical layer functions, rate adaptation and power control, 
in mitigating jamming. 
In the presence of a jammer we find that: 
{\bf (a)} the use of popular rate adaptation algorithms can significantly
degrade network performance and, 
{\bf (b)} appropriate tuning of the carrier sensing threshold 
allows a transmitter to send packets even when being jammed and enables a receiver {\em capture} 
the desired signal. 
Based on our findings, we build ARES, an Anti-jamming REinforcement System, 
which tunes the parameters of rate adaptation and power control 
to improve the performance in the presence of jammers. 
ARES ensures that operations under benign conditions are unaffected. 
To demonstrate the effectiveness and generality of ARES, we evaluate it in three  
wireless testbeds: 
{\bf (a)}  an 802.11n WLAN with MIMO nodes, 
{\bf (b)}  an 802.11a/g mesh network with mobile jammers 
and 
{\bf (c)} an 802.11a WLAN. 
We observe that ARES improves the network throughput across all testbeds by up to 150\%. 
\end{abstract}



\section{Introduction}
\label{sec:intro}
\setcounter{paragraph}{0}

The widespread proliferation of  802.11 wireless networks makes them an attractive target for saboteurs with jamming devices \cite{sesp}.  
Numerous jamming attacks have 
been reported in the recent past \cite{conf-jam, xbox-jam, rf-jam}; 
this makes the defense against such attacks very critical. 
A jammer transmits either dummy packets or simply electromagnetic energy to hinder legitimate communications on the wireless medium.
A jamming attack can 
cause the following effects in an 802.11 network:
	{\bf (a)} Due to carrier sensing, co-channel  transmitters detect activity on the medium and thus, defer their packet transmissions for prolonged periods. 
	{\bf (b)} The jamming signal collides with legitimate packets at receivers. 
As a consequence, the  throughput  is significantly reduced because of these effects. 
Frequency hopping techniques 
have been 
previously proposed 
for avoiding 
jammers \cite{navda07} \cite{Xu04}. 
Such schemes however, are not effective in scenarios with wide-band jammers \cite{wide-jam, intech-jam}. Furthermore,
given that 802.11 operates on relatively few frequency channels, multiple jamming devices operating on 
different channels can significantly hurt performance in spite of using frequency hopping \cite{kpele-wiopt09}. 
More than that, although Frequency Hopping Spread Spectrum was available in the initial 802.11 standard, it was not later included in the 802.11a/b/g standards that are popular today \cite{ieee80211}.

In this paper, we ask the question: 
{\em 
How can legacy 802.11 devices alleviate the effects of a jammer that resides on the same channel as a legitimate communicating pair, in real time? 
} 
We address this challenge by developing  
ARES\footnote{ARES [pron. ``\'aris"] was the god of war in Greek mythology; we choose the name as a symbol of the combat with jammers.}, 
a novel, measurement driven system, which 
detects the presence of jammers 
and invokes rate adaptation and power control strategies to alleviate jamming effects. 
Clearly, not much can be done to mitigate jammers with unlimited resources in terms of transmission power and spectrum efficiency.  
%
Note however that in a plurality of cases the jamming device can be resource constrained, with capabilities similar to that of the legitimate device\footnote{We implement a jamming utility on a commodity 802.11 NIC as described in more detail in Section \ref{sec:experimental}.}.
Portable, battery-operated jammers are typically configured to 
transmit intermittently and sometimes at low power, in order to conserve energy and harm the network for extended periods of time. 
Additionaly, misconfiguration of  ``legitimate" devices can transform them to a resource-constrained jammer \cite{xbox-jam}.  
In these and similar cases, ARES can effectively {\em fight} against the malicious entity, as we discuss later. 
%
%

Our contributions in this paper are the following: 

\noindent {\bf 1. Understanding the impact of jammers in an 802.11 network with rate/power control.}
First, we perform an in-depth measurement-based experimental study on our indoor testbed, to quantify the impact of jamming when employing rate and/or power control.  
To the best of our knowledge,  
there are no such studies to date. 
With rate control, a transmitter can increase or lower its transmission rate depending on the observed 
packet delivery ratio (PDR) at the receiver.
With power control, nodes may increase their transmission powers  
and/or clear channel assessment (CCA) thresholds \cite{powInfocom}
in order to increase the probability of successful packet reception.
The design of ARES is driven by our two key experimental observations: 

\textbf{\em  i) Rate adaptation can be counter-productive:}
In the presence of a jammer that is active intermittently (and sleeps in between), the use of rate adaptation is not always beneficial.
We conduct experiments with three popular rate adaptation algorithms: 
SampleRate \cite{bicket}, Onoe \cite{onoe} and AMRR (Adaptive Multi Rate Retry) \cite{amrr}. 
With every scheme, we observe that {\em the use of rate adaptation may work in favor of the jammer!} 
This is because, rate adaptation wastes a large portion of a jammer's sleeping time in order to gradually converge to the ``best" rate.  
We analytically determine when fixed rate operations may be preferable to the use of rate adaptation. 


\textbf{\em ii) Tuning the carrier sense threshold is beneficial:} 
We collect throughput measurements
with many different transmission powers and CCA thresholds. 
We find that: 
{\bf (a)} In the presence of a jammer, legitimate transmissions with maximum power could lead to
significant benefits, only when operating at low data rates. 
{\bf (b)} Increasing the CCA threshold can allow a transmitter that is being jammed to send packets and at the same time, facilitate the {\em capture} of packets in the presence
of jamming interference; together, these effects can significantly reduce the throughput degradation.  

\noindent {\bf 2. Designing ARES, a novel anti-jamming system.}
The above observations drive the design of ARES. 
ARES primarily  consists of two modules.
The {\bf \em rate control module} decides between fixed-rate assignment and rate adaptation, based on channel conditions and the jammer characteristics.  
The primary objective of this module is to effectively utilize the periods when a jammer is asleep.
The {\bf \em power control module} adjusts the CCA threshold to facilitate the transmission and the reception 
({\em capture}) of legitimate packets during jamming. 
Care is taken to avoid starvation of nodes due to the creation of 
asymmetric links \cite{powInfocom}. 
This module is used to facilitate successful communications while the jammer is active.
Although rate and power control have been proposed as interference alleviation techniques, 
their behavior has not been studied in jamming environments. 
Our work is the first to conduct such a study, as discussed later. 

\noindent {\bf 3. Implementing and experimentally validating ARES.} 
We implement and evaluate the modules of ARES on real hardware,   
thereby making ARES one of the few anti-jamming system implementations for 802.11 networks.  
ARES also contains a jammer detection module that incorporates a mechanism proposed previously in \cite{Xu05}. 
To demonstrate the effectiveness and generality of our system, 
we 
apply it on three different experimental networks:  
a static 802.11n WLAN with MIMO enabled nodes, 
an 802.11a/g mesh network with mobile jammers, and 
a static 802.11a WLAN with uplink TCP traffic. 
Our measurements 
demonstrate that ARES provides 
performance benefits 
in all the three networks; throughput improvements of up to 150\% are observed.

The remainder of the paper is structured as follows.  
In section \ref{sec:background}, we provide some background on jamming and discuss related studies. 
In section \ref{sec:experimental}, we describe our wireless testbed and the experimental methodology. 
We describe our extensive experiments to understand the impact of rate and power control in the presence of a jammer in section \ref{sec:rate}. 
In section \ref{sec:system}, we construct ARES based on our observations. We present our evaluations of ARES in 
section \ref{sec:mimo}. 
Section \ref{sec:discussion} discusses the scope of our study. 
We conclude in section \ref{sec:conclusions}.

\section{Background and Related Work} 
\label{sec:background}
\setcounter{paragraph}{0}

In this section, first we briefly describe the operations of a jammer and its attack capabilities. 
Next, we discuss relevant previous studies.

{\bf Types of Jamming Attacks.} 
Jammers can be distinguished in terms of their attack strategy; a detailed discussion can be found in \cite{Xu05}.  

\textbf{\em Non-stop jamming:} 
{\em Constant} jammers continuously emit electromagnetic energy on a channel. 
Nowadays,  
constant jammers are commercially available and easy to obtain 
\cite{sesp, wide-jam}.  
While constant jammers emit non-decipherable messages, 
{\em deceptive} jammers transmit seemingly legitimate back-to-back dummy data packets. 
Hence, they can mislead other nodes and monitoring systems into believing that legitimate traffic is 
being sent. 

\textbf{\em Intermittent Jamming:} 
As the name suggests, these jammers are active intermittently; the primary goal is to conserve battery life. 
A {\em random} jammer typically alternates between uniformly-distributed jamming and sleeping periods; 
it jams for $T_{j}$ seconds and then it sleeps for $T_{s}$ seconds.  
A {\em reactive} jammer 
starts emitting energy only if it detects traffic on the medium.  This makes the jammer difficult to detect. However, implementing reactive jammers can be a challenge. 

For the purposes of this work,
we primarily consider the random 
jammer
model. 
Attackers are motivated into using a random jammer because putting the jammer to sleep intermittently can increase its lifetime and decrease the probability of detection \cite{Xu05}.  
Furthermore, it is the most generalized representation of a jammer; 
appropriately choosing the sleep times could
turn the jammer into a constant jammer or (with high probability) a reactive jammer.  
Moreover, reactive jammers  
are not easily available since they are harder to implement and require special expertise on the part of the attacker. 
We discuss the applicability of ARES with constant and reactive jammers, in section \ref{sec:discussion}.

{\bf Related work.}  
Most previous studies employ frequency hopping to avoid jammers. 
Frequency hopping, however, cannot 
alleviate the influence of a wide-band jammer \cite{wide-jam, intech-jam}, which can effectively jam all the available channels. 
In addition, recent studies have shown that a few cleverly-coordinated, narrow-band jammers can practically block the whole spectrum \cite{kpele-wiopt09}. 
Thus, 
ARES does not rely on frequency hopping.    
%
%
%

\textbf{\em Studies based on frequency hopping:} 
Navda \textit{et al.} \cite{navda07} implement a proactive frequency hopping protocol with pseudo-random channel switching. 
They compute  
the optimal frequency hopping parameters, assuming that the jammer is aware of the procedure followed. 
Xu  \textit{et al.} \cite{Xu04} propose two anti-jamming techniques: reactive channel surfing 
and spatial retreats. 
However, their work is on sensor networks which only support very low data rates and transmission powers. 
Gummadi \textit{et al.} \cite{Gummadi07} find that 802.11 devices are vulnerable to specific patterns of narrow-band interference  related to time recovery, dynamic range selection and PLCP-header processing.   
They show that due to these limitations, an intelligent jammer with a 1000 times weaker signal (than that of the legitimate transceiver) can still corrupt the reception of packets. 
In order to alleviate these effects, they propose a rapid frequency hopping strategy. 
%

\textbf{\em Other relevant work:} 
Xu \textit{et al.} \cite{Xu05} develop efficient mechanisms for jammer detection at the PHY layer (for all the 4 types of jammers).  
However, they do not propose any jamming mitigation mechanisms. 
In \cite{Xu06}, the same authors suggest that  competition strategies, 
where transceivers adjust their 
transmission powers and/or 
error correction codes, {\em might} alleviate jamming effects. 
However, they neither propose an anti-jamming protocol nor perform evaluations to validate their suggestions. 
Lin and Noubir \cite{Lin03} 
present an analytical evaluation of the use of cryptographic interleavers with different coding schemes to improve the robustness of wireless LANs. 
In \cite{Noubir03}, the authors show that in the absence of error-correction codes (as with 802.11) 
the jammer can conserve battery power by destroying only a portion of a legitimate packet. 
Noubir \cite{Noubir} also proposes  
the use of a combination of directional antennae and node-mobility in order to alleviate jammers. 
ARES can easily be used in conjunction with directional antennae or with error correction codes.

\textbf{\em Prior work on rate and power control:} 
Rate and 
power control techniques 
have been proposed in the literature, as means of mitigating interference (e.g. \cite{wong06, bicket, powInfocom, shah05} and the references therein).  
However, they do not account for a hostile jamming environment;  
with these schemes, 
nodes  cooperate in order to mitigate the impact of "legitimate" interference, 
thereby improving the  performance.  
On the other hand, ARES is specialized towards handling malicious interference of  jammers, 
which attempt to disrupt ongoing communications.


\section{Experimental Setup} 
\label{sec:experimental}
\setcounter{paragraph}{0}

In this section, we describe our wireless testbed and the 
experimental methodology that we follow. 

\begin{figure}[t]
\begin{center}
\includegraphics[width=5cm]{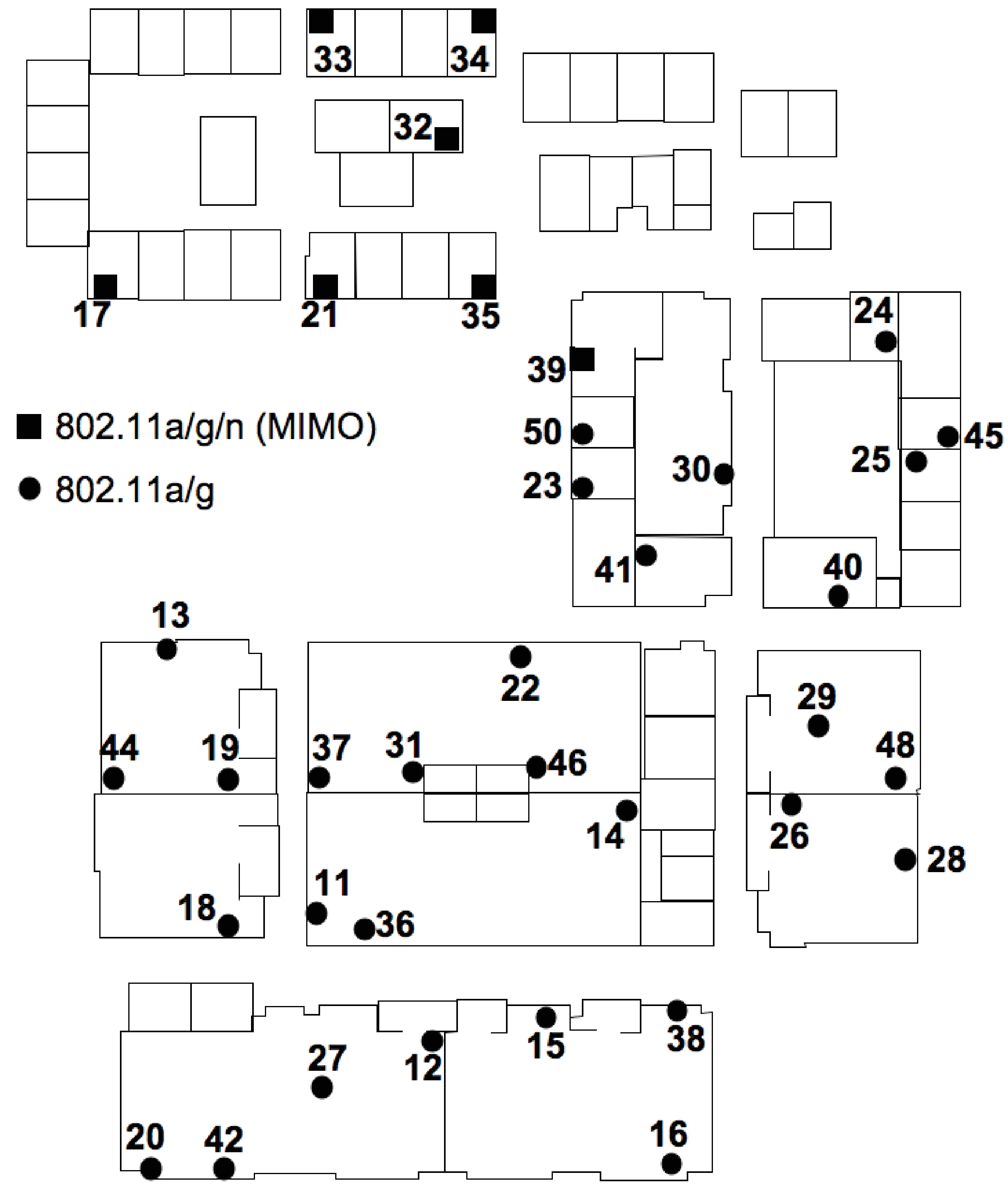}
\caption{The deployment of our wireless testbed.} \label{fig:ucr}
\end{center}
\end{figure}

{\bf Testbed Description:}
Our wireless testbed \cite{trid} is deployed in the third floor of Engineering Building II, 
at the University of California, Riverside.  
Our testbed consists of 37 Soekris net4826 nodes \cite{ucrtestbed}, which mount a Debian Linux distribution with kernel v2.6, over NFS.  
The node layout is depicted in Figure \ref{fig:ucr}.
Thirty of these nodes are each equipped with two miniPCI 802.11a/g WiFi cards, 
an \textit{EMP-8602 6G} with Atheros chipset and an \textit{Intel-2915}. 
The other 7 nodes are equipped with one \textit{EMP-8602 6G} and one \textit{RT2860} card that supports MIMO-based (802.11n) communications. 
We use the MadWifi driver \cite{madwifidriver} for the \textit{EMP-8602 6G} cards. 
We have modified the Linux client driver \cite{ralinkdriver} of the \textit{RT2860} 
 to enable STBC  
(Space Time Block Coding) support. 
We use a proprietary version of the {\em ipw2200} AP (access point) and client driver/firmware 
of the \textit{Intel-2915} card. 
With this version we are able to tune the CCA threshold parameter. 

{\bf Experimental Settings and Methodology:}
We experiment with different rate adaptation algorithms in the presence of random jammers. 
We also perform experiments with various 
transmission powers 
of jammers and powers/CCA thresholds of legitimate nodes. 
Our 
measurements encompass an  
exhaustive set of wireless links, 
routes of different lengths, as well as static and mobile jammers. 
We examine  
both SISO and MIMO links. 
We experiment with three modes of operation: 802.11a/g/n 
 (unless otherwise stated throughout this paper, our observations are consistent for all three modes of operation).
 The experiments are performed late at night in order to isolate the impact of the jammers by avoiding interference from co-located WLANs. 
By default, all devices (legitimate nodes and jammers) set their transmission powers to $18$ dBm. 

\textbf{\em Implementing a random jammer:} 
Our implementation of a random jammer is based on a specific configuration (CCA = 0 dBm) and a user space utility that sends broadcast packets as fast as possible. 
For the purposes of research, we have implemented our own random jammer 
on an 802.11 legacy device, by  
setting the CCA threshold to $0$ dBm. 
By setting the CCA threshold to such a high value,
we force the device to ignore all legitimate 802.11 signals even after carrier sensing; 
packets arrive at the jammer's circuitry with powers less than 0 dBm
(even if the distances between the jammer and the legitimate transceivers are very small). 
An effective random jammer should be able to transmit packets on the medium, as fast as possible, during random {\em active} time intervals. 
We develop a user-space software utility with the following functionalities: 
\begin{itemize} 
\item 
The jammer transmits broadcast UDP traffic. 
This ensures that its packets are transmitted back-to-back and that the jammer does not wait for any ACK messages  (by default the backoff functionality is disabled in 802.11 for broadcast traffic); 
in other words, this set up allows the jamming node to defer its back-to-back transmissions for the minimum possible time (i.e. $DIFS + min_{BackOff}$).  
Our utility employs {\em raw sockets}, which allow the construction of UDP packets from scratch and the forwarding of each packet directly down to the hardware\footnote{Administration privileges are required for this operation.}. 
Note that this implementation {\em bypasses the 802.11 protocol} and hence, the jammer does not wait in the backoff state after each packet transmission. 
\item 
Our utility schedules uniformly-distributed random jamming intervals. 
The jammer is in the active state for a random period of time, during which it constantly transmits packets back-to-back. 
It then transits to an idle (sleeping) state for a different, randomly chosen period of time during which it does not emit energy. 
The two states alternate and their durations 
are computed anew at the beginning of each cycle (a cycle consists of an active and an idle period). 
\end{itemize}
We use a set of 4 nodes as jammers on our testbed; these 
are equipped with  {\em Intel-2915} cards which allow CCA tuning. 

\textbf{\em Traffic characteristics:} 
We utilize the {\em iperf} measurement tool to generate UDP data traffic 
among legitimate nodes; 
the packet size is 1500 bytes. 
The duration of each experiment is 1 hour. 
For each experiment, we first enable {\em iperf} traffic between legitimate nodes, and subsequently, we activate the jammer(s). 
We consider both mesh and WLAN connectivity. 
%
%
%
%
We experiment with different jammer distributions, namely:   
{\bf (a)} \textit{frequent jammers}, which are active almost all of the time, 
{\bf (b)} \textit{rare jammers}, which spend  most of their time sleeping, and 
{\bf (c)} \textit{balanced jammers} that have similar average jamming and sleeping times. 
We have disabled 
RTS/CTS message exchange 
throughout our experiments 
(a common design decision in practice 
\cite{etx}).

\section{Deriving System Guidelines}
\label{sec:rate}
\setcounter{paragraph}{0}

In this section, we describe  our experiments towards understanding the  
behavioral trends of power and rate adaptation techniques, in the presence of random jammer(s). 
Our goal is 
to determine if there are properties that can be exploited 
in order to alleviate jamming effects. 
We perform experiments on both single-hop and multi-hop configurations. 


\subsection{Rate Adaptation in Jamming Environments}
\label{sec:ratemeas} 
Rate adaptation algorithms are utilized to select an appropriate transmission rate as per the current channel conditions. 
As interference levels increase, lower data rates are dynamically 
chosen. 
Since legitimate nodes consider jammers as interferers, rate adaptation will reduce the transmission rate on legitimate links while jammers are active. 
Hence, one could potentially argue that rate control on legitimate links increases reliability by reducing rate and thus, can provide throughput benefits in jamming environments. 

To examine the validity of  
this argument, 
we experiment with three different, popular rate adaptation 
algorithms,
SampleRate \cite{bicket}, AMRR \cite{amrr} and Onoe \cite{onoe}. 
These algorithms are already implemented on the MadWifi driver that we use. 
For simplicity, we first consider a balanced jammer, which selects the sleep duration from a uniform 
distribution $U[1,8]$ 
and the jamming duration from $U[1,5]$ (in seconds).

\textbf{\em Details on the experimental process:} 
We perform experiments with both single-hop and multi-hop configurations. 
For each experiment, we first load the particular rate-control Linux-kernel module (SampleRate, AMRR or Onoe) on the wireless cards of legitimate nodes. 
We initiate data traffic between the nodes and after a random time, we activate the jammer. 
We collect throughput measurements on each data link once every 500 msec.  
We use the following terminology: 

%
{\em 1) Fixed transmission rate $R_f$:} This is the nominal transmission rate configured on the wireless card.  

{\em 2) Saturated rate $R_s$:} It is the rate achieved when $R_f$ is chosen to be the rate on the wireless card.  
In order to compute $R_s$, for a given $R_f$, we consider links where the packet delivery ratio (PDR)\footnote{We refer to the application layer packet delivery ratio, which includes the MAC layer retransmissions.} is 100 \% for the particular setting of $R_f$; we then
measure the rate achieved in practice. We notice that for lower values of $R_f$, the specified rate is actually achieved on such links. However, for
higher values of $R_f$ (as an example $R_f = 54$ Mbps), the achieved data rate is much lower due to MAC layer overheads, such as MAC layer retransmissions \cite{atheros-paper}.
Table \ref{tab:saturated} contains a mapping, derived from measurements on our testbed, between $R_f$ and $R_s$.  

{\em 3) Application data rate $R_a$:} This is the rate at which the application generates data.  
%

\begin{table}[h]
\centering \small
\begin{tabular}{|c|c|c|c|c|c|c|c|c|c| l|r|}
\hline
$R_f$ & 6 & 9 & 12 & 18 & 24 & 36 & 48 & 54\\	
\hline
 $R_s$ & 6 & 9 & 12 & 18 & 24 & 26 & 27 & 27\\
\hline
\end{tabular} 
\caption{The saturated-throughput matrix in Mbps.}
\label{tab:saturated}
\end{table}
It is difficult (if not impossible) to a priori determine the {\em best} fixed rate on a link. Given this difficulty, we set $R_f = \{min~R_f:~R_f\ge R_a\}$, which is the maximum rate that is required by the application (we discuss the implications
of  this
choice later).  
%
Our key observations 
are summarized below: 

\begin{itemize}
\item 
\textbf{Rate adaptation algorithms perform poorly on  
high-quality   
 links due to the long times that they incur for converging to the appropriate high rate.}  
\item 
\textbf{On {\em lossless} links, the fixed rate $R_f$ is better, while rate adaptation is beneficial on {\em lossy} links. }
\end{itemize}
We defer defining what constitute  lossless or lossy links to later; 
conceptually, we consider lossless links to be those links that can achieve higher long-term throughputs using a fixed transmission rate $R_f$ 
, rather than by applying rate adaptation. 

\subsubsection{Single-hop Configurations}
\label{sec:singleho}

Our experiments with one-hop connectivity involve $80$ sets of sender-receiver pairs and one jammer per pair. 
We impose that a jammer interferes with {\em one} link at a time and that the legitimate data links do not interfere with each other. 
Thus, we perform 20 different sets of experiments, with 4 isolated data links and 4 jammers in each experiment.  

\begin{figure*}[ht]
\begin{center} 
\parbox{4.5in} { 
     \centerline{  
			\includegraphics[scale=0.2,angle=270]{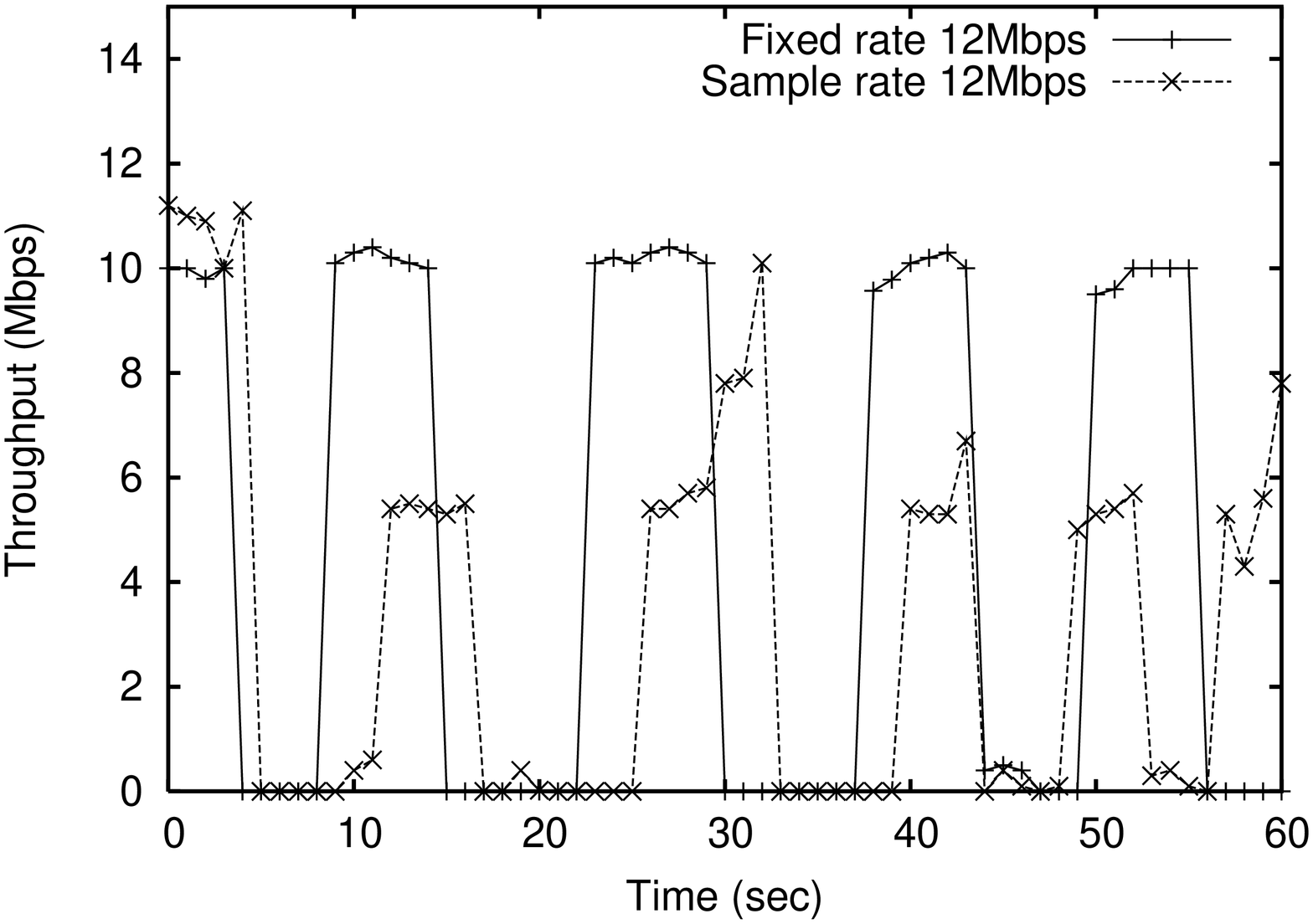} 
			\includegraphics[scale=0.2,angle=270]{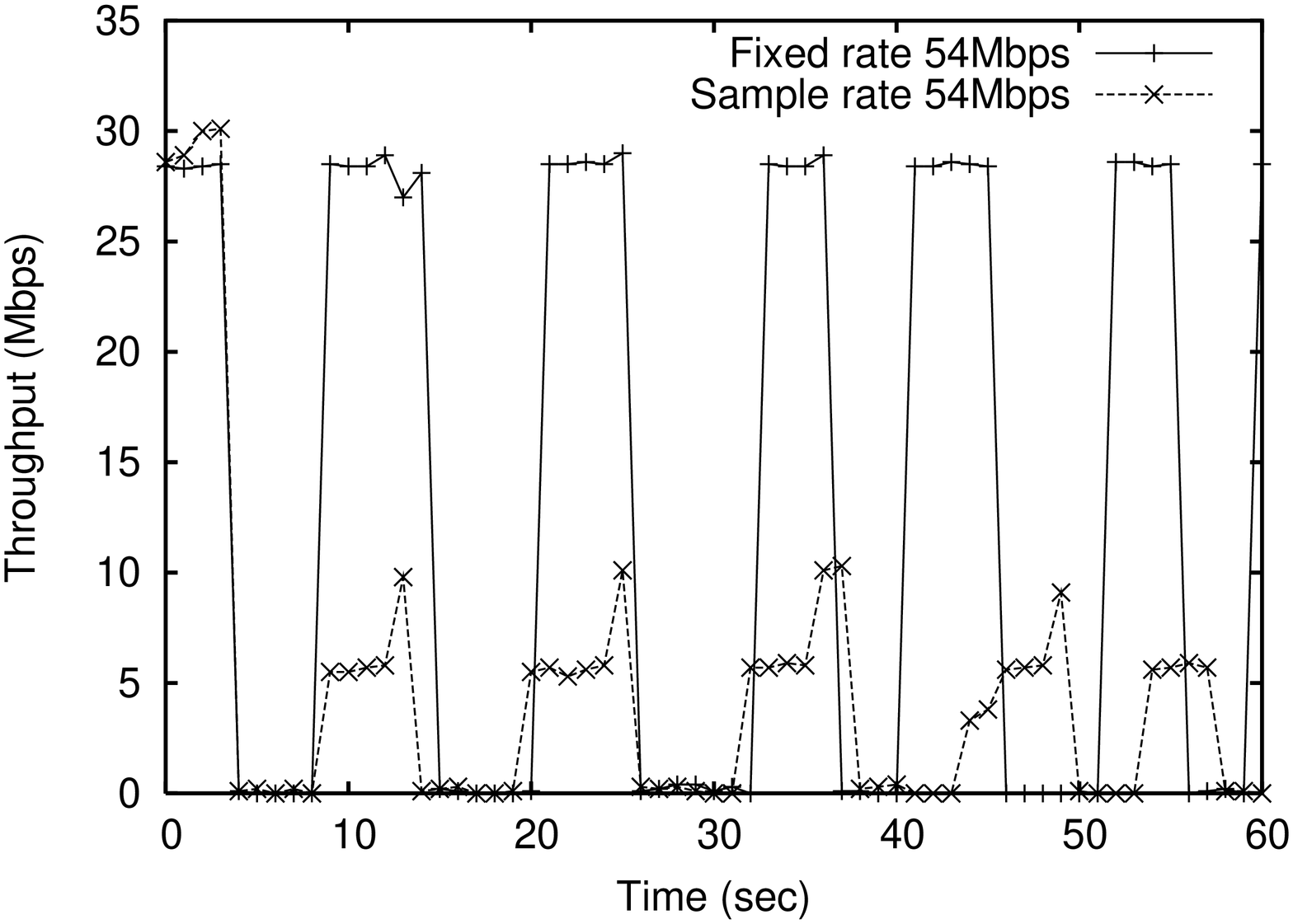}
	} 
    \caption{Rate adaptation algorithms may not find the best rate during the sleep period of the jammer. 
				We show cases for 2 different links, one with $R_a=12$ Mbps (left) and one with $R_a=54$ Mbps (right). }
    \label{fig:time_series}
} 
\makebox[0.1in] {}
\parbox{2.3in} {
     \centerline{\includegraphics[scale=0.2,angle=270]{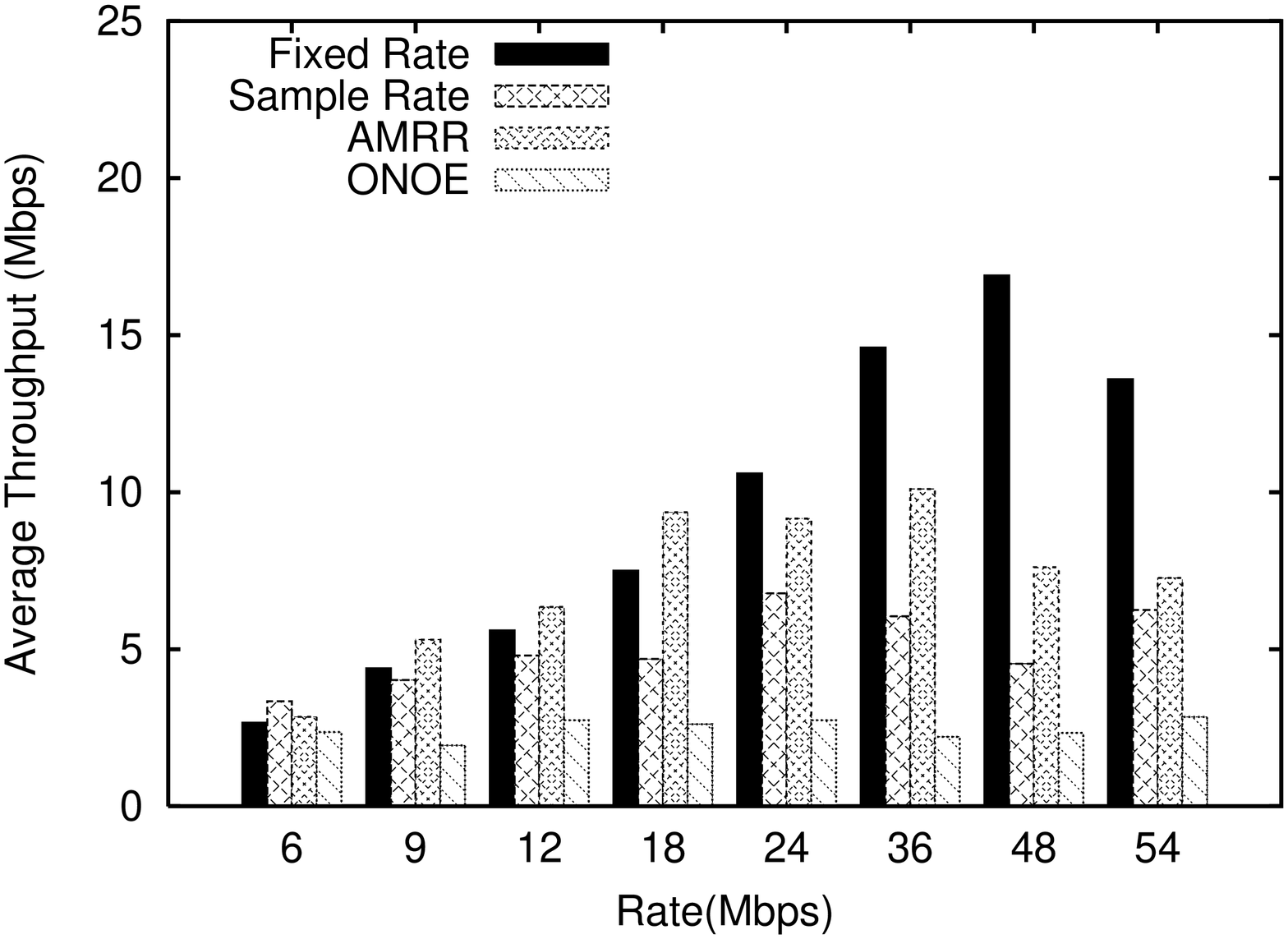}}  
     \caption{Fixed rates outperform rate adaptation for high-quality links, under random jamming. ($R_a=R_f$)}
     \label{fig:fig9}
}
\end{center}
\end{figure*}

\textbf{Rate adaptation consumes a significant part of the jammer's sleep time, to converge to the appropriate rate:} 
As soon as the jammer ``goes to sleep", the link quality improves and thus, 
the rate control algorithm starts increasing the rate progressively.  
However, since the purpose of a jamming attack is to corrupt as many transmissions as possible, 
the jammer will typically not sleep for a long time. 
In such a case, the sleep duration of the jammer will not be enough for the rate control to reach the highest rate possible.
To illustrate this we choose two links on our testbed, one that can support 12 Mbps and the other that can support 54 Mbps.
Figure \ref{fig:time_series} depicts the results. 
We observe that 
	{\bf (a)} irrespective of whether SampleRate or a fixed rate strategy is used,  during jamming 
the throughput drops to values close to zero since the jammer blocks the medium for the sender, and 
	{\bf (b)} {\em the throughput achieved with SampleRate is quite low, and much lower than if we fix the rate to the constant value of 12 Mbps.} 
Note that we have observed the same behavior with AMRR and Onoe. 

%
%

\textbf{Fixed rate assignment outperforms rate adaptation on lossless links:} 
As alluded to above, in order to find the {\em best} rate on a link after the impact of a jammer, 
the rate adaptation mechanisms gradually increase the rate, invoking transmissions at all 
the lower rates interim, 
 until the best rate is reached. 
For links that can inherently support high rates, this process  
might consume the sleep period of the jammer 
(as suggested by the results in Figure \ref{fig:time_series}). 
If the best rate for a link was known a priori, 
at the instance that the jammer goes to sleep, transmissions may be invoked at that rate. 
This would utilize the sleep period of the jammer more effectively. 
As observed in Figure \ref{fig:fig9}, the throughputs achieved with fixed rate assignment are much higher than those achieved with rate adaptation on such links. 
\vspace{0.1cm}

\noindent{\bf \em Determining the right transmission rate policy:} 

\vspace{0.1cm}
{\em Implications of setting $R_f = \{min~R_f:~R_f\ge R_a\}$:} 
Since the application
does not require the link to sustain a higher rate, the highest throughput for that application rate is reached either with 
this choice of $R_f$ or with some rate
that is lower than $R_a$. 
If the rate adaptation algorithm converges to a rate that results in a throughput that is higher than with 
 the chosen $R_f$, then the adaptive rate strategy
should be used. If instead, during the jammer's sleep period, the rate adaptation technique is unable to converge to such a rate, the fixed rate strategy is better.

{\em Analytically determining the right rate:} In order to determine whether it is better to use a fixed or an adaptive-rate approach for a given link, we perform an analysis 
based on the following parameters: 

\begin{enumerate}
\item The distribution of the jammer's active and sleep periods (we call this the {\em jammer's distribution}). 
\item The application data rate, $R_a$. 
\item The performance metric on the considered legitimate link, i.e., PDR, link throughput, etc.
\item The rate adaptation scheme that is employed, i.e., Onoe, SampleRate, etc. 
The key scheme-specific factor is the transition time from a lower rate to the next higher rate, under conducive conditions. 
\item The {\em effectiveness} of the jammer $F$, measured by the achievable throughput while the jammer is on. The lower the throughput, the more effective the jammer. 
\end{enumerate}

Let us suppose that the expected {\em sleeping} duration of the jammer during a cycle, is given by $E[t_{s}]$ and the expected period for which 
it is active, by $E[t_{j}]$. The expected duration of a {\em cycle} is then $ E[t_{s}] + E[t_{j}]$. 
As an example, if the jammer picks its sleeping period from a 
uniform distribution $U[a,b]$  and its jamming period from $U[c,d]$, $E[t_{s}]$ and $E[t_{j}]$ are equal to $\frac{b+a}{2}$ and $\frac{d+c}{2}$, respectively. 
For simplicity let us assume that the link-quality metric 
employed\footnote{Our analysis can be 
modified to adopt any other link-quality metric.} 
is the PDR. 
With application data rate $R_a$ and \textbf{{\em fixed}} transmission rate $R_f$, 
the throughput achieved during a jammer's cycle is: 
%
\begin{equation}
T_{fixed} = \frac{E[t_{s}]}{E[t_{s}]+E[t_{j}]}\cdot PDR_f \cdot R_s +  \frac{E[t_{j}]}{E[t_{s}]+E[t_{j}]}\cdot F,
\label{eq:fixedtput}
\end{equation}
where $PDR_f$ is the PDR of the link at rate $R_f$.  
Recall that the rate achieved in practice with a specified rate $R_f$ is $R_s$.  

\noindent To compute the throughput with \textbf{\em rate adaptation}, we proceed as follows.  
Let us assume that $x(F, R_s)$ 
corresponds to the convergence time of the rate adaptation algorithm (specific to the chosen algorithm). 
We consider the following two cases.

\noindent {\bf 1)} {\boldmath $x(F, R_s)< E[t_{s}]$}.  
This case holds when the jammer's sleep duration is sufficient (on average) for the rate control algorithm to 
converge to the best rate $R_s$. 
In this scenario, the achievable throughput is:
\begin{equation*}
T_{adapt} = \frac{\left[E[t_{s}]-x(R_s)\right]\hspace{-0.1cm}\cdot \hspace{-0.1cm}R_s + \displaystyle\sum_{R_{i}} y(R_{i})\hspace{-0.1cm}\cdot\hspace{-0.1cm} R_{i} + E[t_{j}]\cdot F}{E[t_{s}]+E[t_{j}]}, 
\label{eq:adapt1}
\end{equation*}
where $R_{i} \in S$, $S$ being the set of all intermediate rates from $F$ to $R_s$.  
$y(R_{i})$ is the time that the rate control algorithm spends at the corresponding rate $R_{i}$. 
The values of $y(R_{i})$  
are specific to the implementation of the rate control algorithm.  
Note that $x(F, R_s)$ can be easily computed from  $y(R_i)$ by adding all the individual durations 
for the rates belonging to the set $S$. 

\noindent {\bf 2)} {\boldmath $x(F, R_s)\geq E[t_{s}]$}. 
In this scenario, the average sleep time of the jammer is insufficient for the rate control algorithm to converge to the desired rate. 
When the jammer wakes up, the rate will again drop to lower levels due to increased interference. 
Here, the throughput that can be achieved during a jammer's cycle is: 
%
%
%
%
\begin{equation*}
T_{adapt} = \frac{\displaystyle\sum_{i=1}^n y(R_{i})\hspace{-1mm}\cdot\hspace{-1mm} R_{i} \hspace{-1mm}+\hspace{-1mm} 
\left[E[t_{s}] \hspace{-1mm}-\hspace{-1mm} \displaystyle\sum_{i=1}^n y(R_{i})\right]\hspace{-1mm}\cdot\hspace{-1mm} R_{n+1} + 
E[t_{j}]\hspace{-0.5mm}\cdot \hspace{-0.5mm} F}{E[t_{s}]+E[t_{j}]} 
\label{eq:adapt2}
\end{equation*}
%
 where $n=max\{k:\displaystyle\sum_{i=1}^k y(R_{i})\leq E[t_{s}]$ \}. 

Based on the above analysis, we define a link to be {\bf lossy}, when $T_{fixed} \leq  T_{adapt}$; the links on which $T_{fixed} >  T_{adapt}$ 
are classified as {\bf lossless} links.  
Clearly for lossy links it is better to use the rate adaptation algorithm. 
The analysis can be used to compute $PDR_{f}^{TH}$, a threshold value of $PDR_f$  
 below which, 
a rate adaptation strategy performs better than the fixed rate approach. 
In particular, by setting  $T_{fixed} = T_{adapt}$ and solving this equation,
one can compute $PDR_{f}^{TH}$. 
Based on this, a decision can be made on whether to enable rate adaptation or use fixed-rate assignment. 
If the observed PDR is larger than the computed threshold, fixed rate should be used; otherwise, rate adaptation should be used.  

\begin{figure*}[ht]
\begin{center}
\parbox{2in} {
     \centerline{ \includegraphics[scale=0.4]{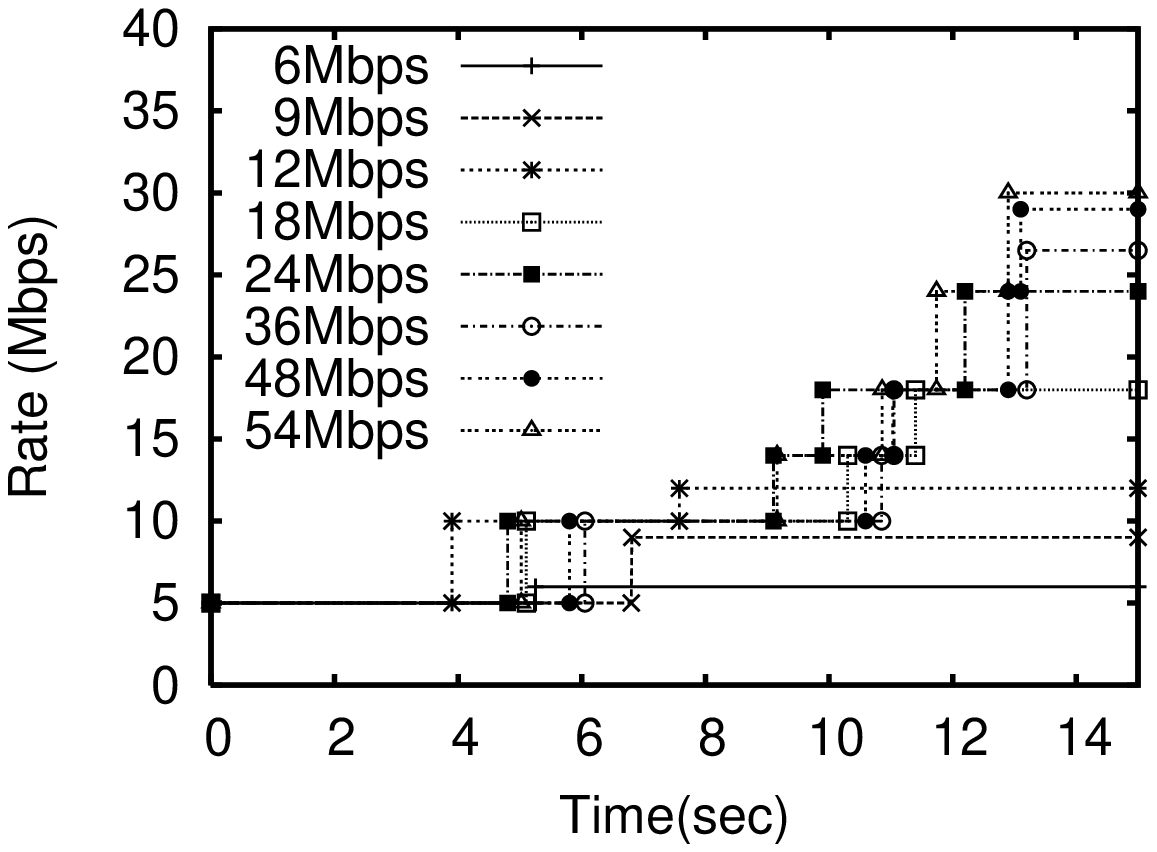}}\vspace{-0.15in}
\hspace{-1in}
\vspace{-0.1in}
     \caption{Measured convergence times of the MadWifi SampleRate algorithm, for the different application data rates.}\label{fig:fig8}
}\hspace{0.5cm}
\makebox[.06in] {}
\parbox{2in} {
     \centerline{\includegraphics[scale=0.4]{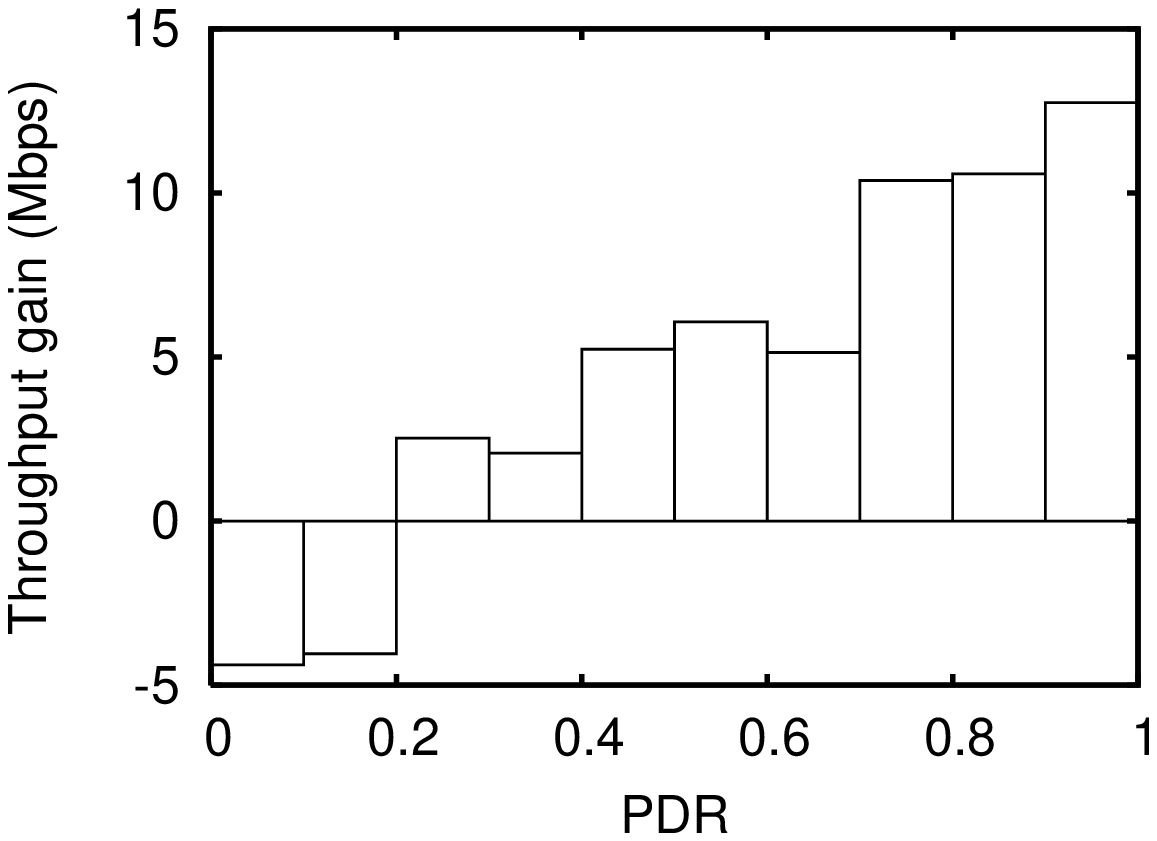}}
\vspace{-0.1in}
     \caption{Throughput gain of fixed rate Vs. SampleRate, for various link qualities and for  application data rate of  54 Mbps.}\label{fig:tputadv}
}\hspace{0.5cm}
\makebox[.06in] {}
\parbox{2in} {
     \centerline{\includegraphics[scale=0.2,angle=270]{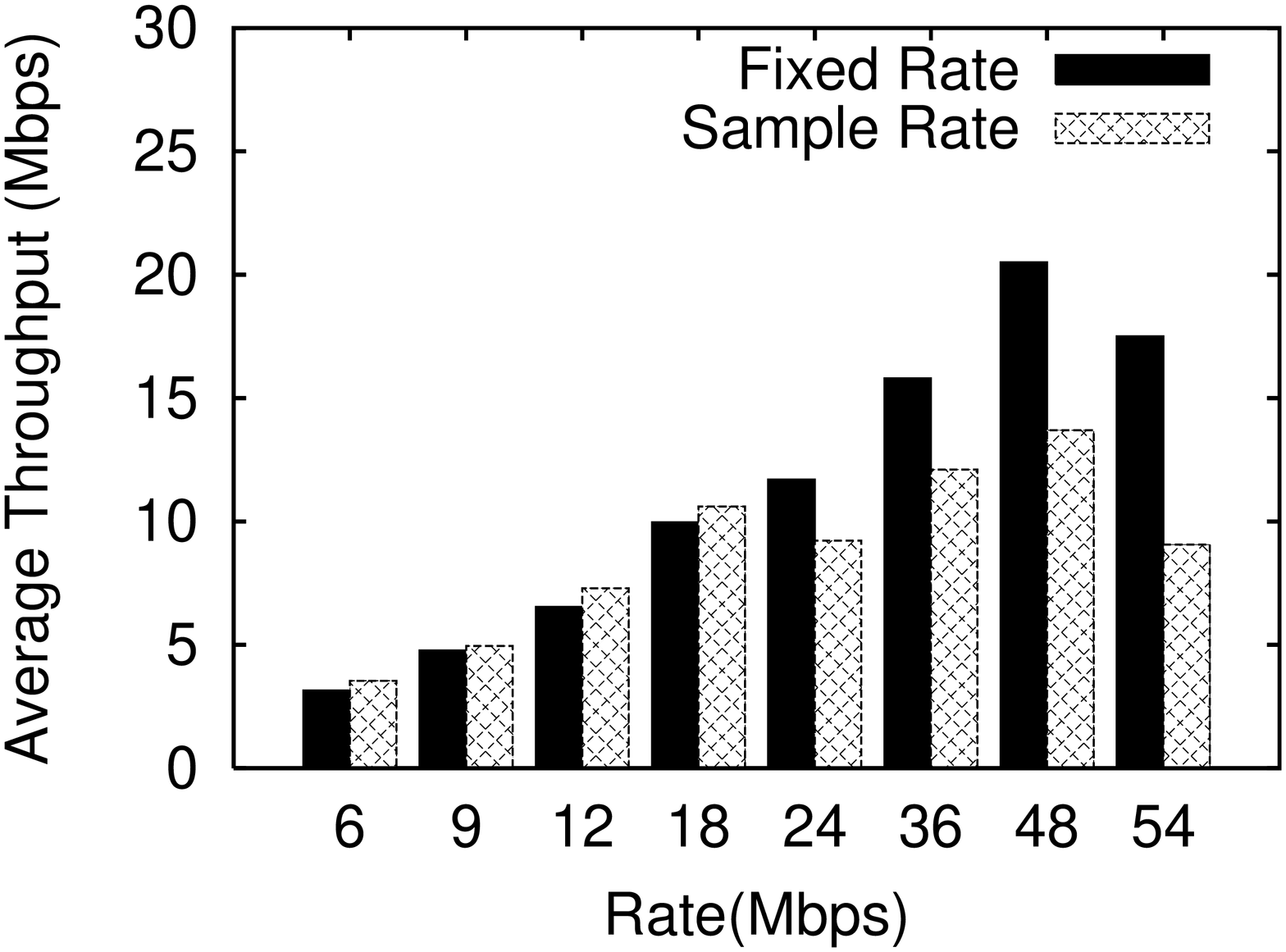}}
\vspace{-0.1in}
     \caption{The performance with rare jammers is aligned with our observations for the case with balanced jammers. ($R_a=R_f$)}\label{fig:rarejam}
}
\end{center}
\end{figure*}


\textbf{Validation of our analysis: }
In order to validate our analysis, we measure $PDR_{f}^{TH}$ 
on $80$ different links in the presence  of a balanced jammer. 
We then compare them against the $PDR_{f}^{TH}$ values computed with our analysis. 
Note here that the analysis itself depends on measured values of certain quantities 
(such as the jammer distribution and the function $y(R_i)$). 
In this experiment, we consider the SampleRate algorithm, and measure the values of $x(F, R_s)$ and $y(R_i)$. 
The jammer's sleep time follows  $U[0,4]$ and the jamming time follows $U[1,6]$. 
Figure \ref{fig:fig8} plots the values of function $y$ for different values of $R_f$. 

%

In Table \ref{table:PDRthr}, we compare the theoretically computed PDR thresholds with the ones measured on our testbed, for various values of $R_f$. 
We observe that the $PDR_f$ thresholds computed with our analysis are very similar to the ones measured on our testbed.  
There are slight discrepancies since our analysis is based on using measured average values which may change to some extent over time.
We wish to stress that while we verify our analysis assuming that the jammer is active and idle for uniformly distributed periods of time, our analysis depends only on expected values and is therefore valid for other jammer distributions.
Finally, Figure \ref{fig:tputadv} shows the advantage of using a fixed rate approach over SampleRate for various PDR values and with 
$R_f$ = 54 Mbps.  
We observe that SampleRate provides higher throughputs only for very low PDR values. 

\begin{table}[ht]
\centering \small
\begin{tabular}{|c|c|c|}
\hline $R_f$ & Measured $PDR_{f}^{TH}$ & Analytical $PDR_{f}^{TH}$ \\ 
\hline 6 & 0.82 & 0.83 \\ 
\hline 9 & 0.52 & 0.55 \\ 
\hline 12 & 0.40 & 0.41 \\ 
\hline 18 & 0.26 & 0.27 \\ 
\hline 24 & 0.19 & 0.21 \\ 
\hline 36 & 0.19 & 0.20 \\ 
\hline 48  & 0.17 & 0.185 \\ 
\hline 54 & 0.15 & 0.185 \\ 
\hline 
\end{tabular} 
\caption{$PDR_f$ thresholds}
\label{table:PDRthr}
\end{table}

\begin{figure*}[th]
\begin{center}
\parbox{2in} {
     \centerline{ \includegraphics[width=3.4cm,angle=270]{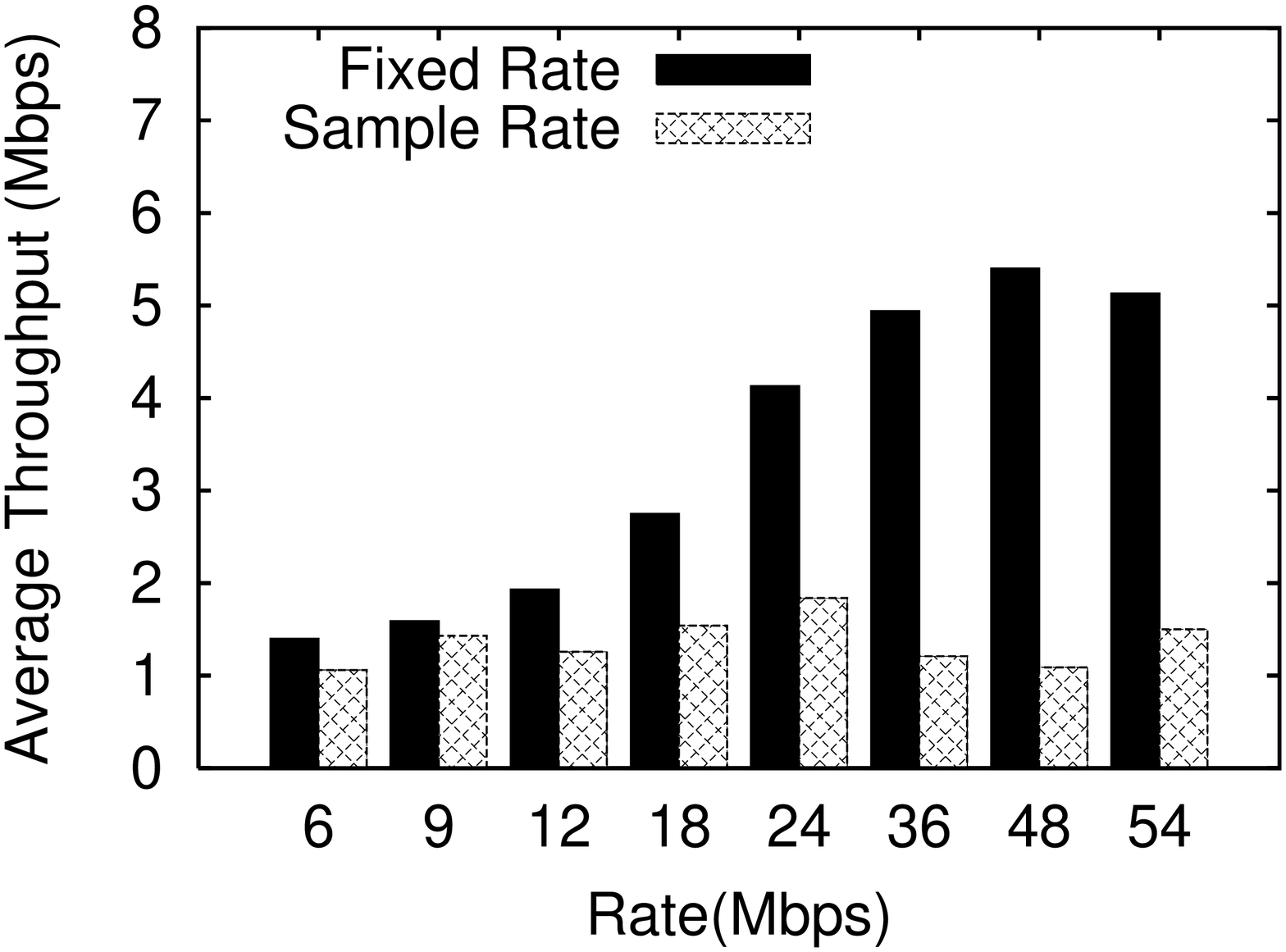}}
     \caption{Fixed rate improves the performance more than rate adaptation at high rates, with frequent jammers. ($R_a=R_f$)}\label{fig:freqjam}
}
\makebox[.3in] {}
\parbox{2in} {
     \centerline{\includegraphics[width=3.4cm,angle=270]{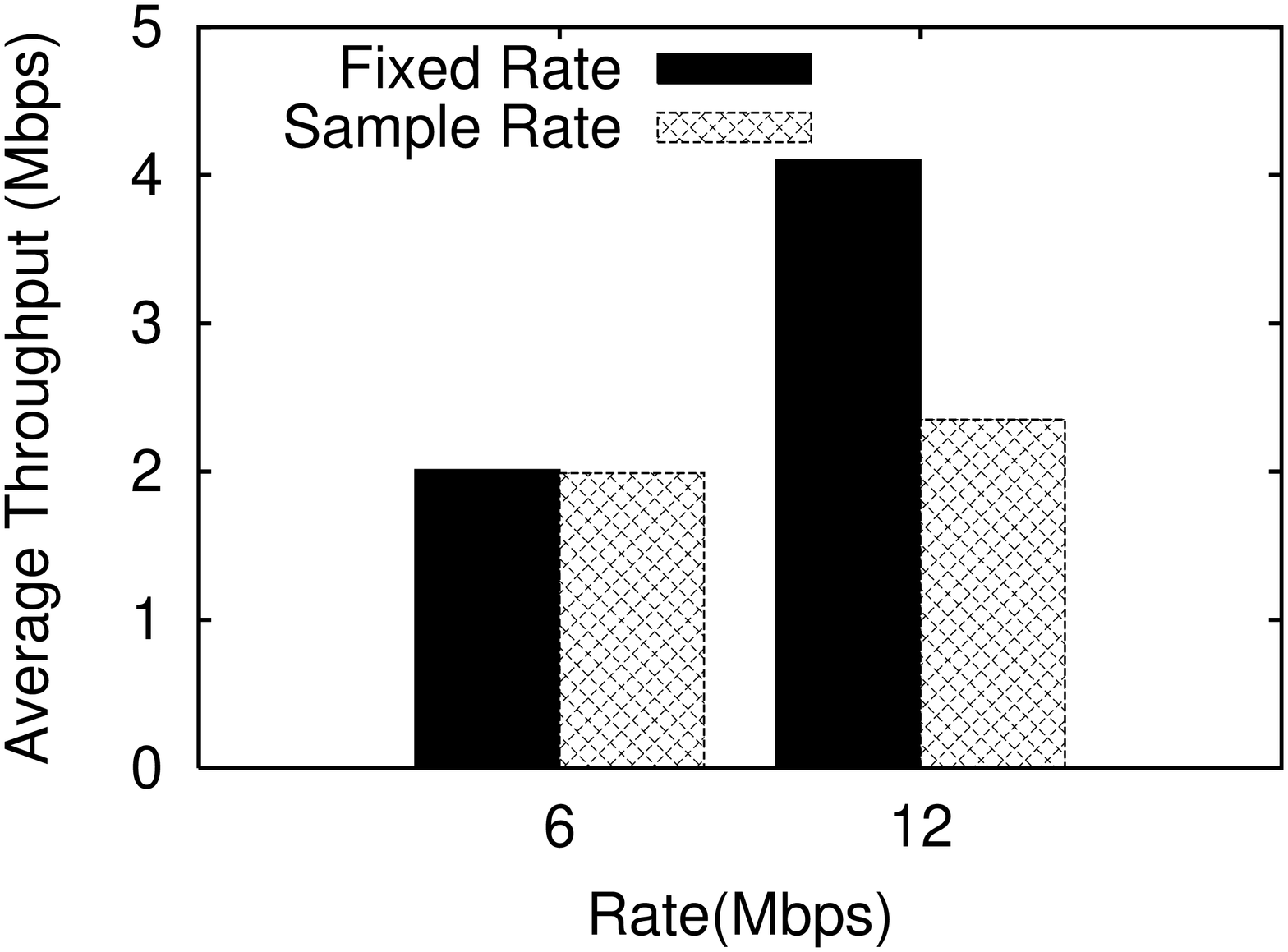}}
     \caption{Rate adaptation presents the same behavior in multihop links; it provides lower throughput at high rates.}\label{fig:route}
}
\makebox[.3in] {}
\parbox{2in} {
     \centerline{\includegraphics[width=5cm]{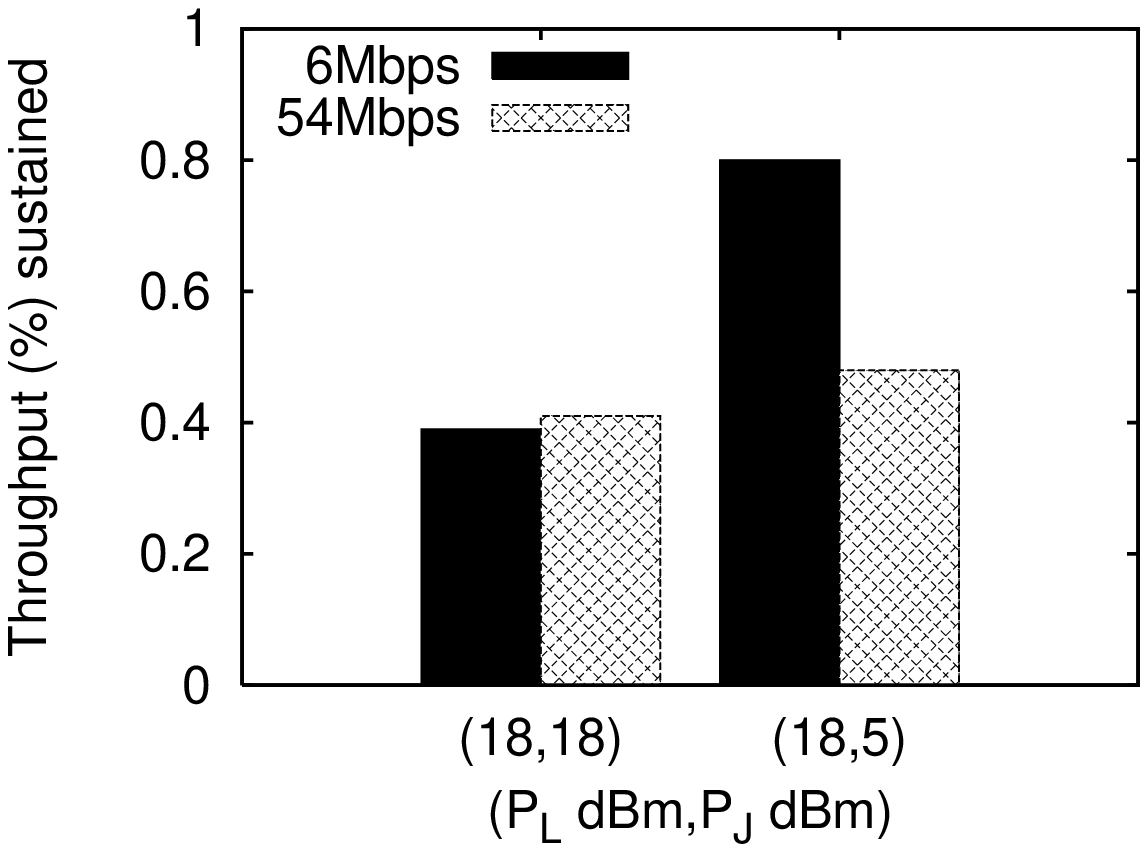}}\vspace{-0.1in}
     \caption{Percentage of the isolated throughput, for various $P_L$ and $P_J$ combinations, for two different transmission rates.}\label{fig:rssi_bars}
}
\end{center}
\end{figure*}

Next, we consider two extreme cases of jamming: frequent and rare jammers (see section \ref{sec:experimental}).  
The distributions that we use in our experiments for these jammers are shown in Table \ref{tab:jdistro}.  
Note that by choosing the jammer's sleeping and jamming time from distributions like the one of the frequent jammer, we essentially construct a constant jammer.  
With frequent jammers, the difference in the performance between fixed rate assignment and rate adaptation is larger, while for a rare jammer it is smaller.  
This is because with rare jamming, rate adaptation will have more time to converge and therefore often succeeds in achieving 
the highest rate possible; 
one observes the opposite effect when we have a frequent jammer.  
The results are plotted in Figures \ref{fig:rarejam} and \ref{fig:freqjam}.  

\begin{table}[ht]
\vspace{-0.02in}
\centering \small
\begin{tabular}{|c|c|c|}
\hline - & Sleep time (sec)  & Jamming time (sec) \\ 
\hline Balanced & U[1,8] & U[1,5] \\ 
\hline Rare & U[1,5] & U[1,2] \\ 
\hline Frequent & U[1,2] & U[1,15] \\ 
\hline
\end{tabular} 
\caption{The jamming distributions that we use in our experiments. }
\label{tab:jdistro}
\end{table}

\subsubsection{Random Jamming in Multi-hop Topologies}
\label{sec:multihoprate}
Next, we examine the impact of a random jammer on the end-to-end throughput of a multi-hop path. 
We experiment with 15 different routes on our testbed. 
We fix static routes of various lengths (from 2 to 4 links per route) utilizing the \textit{route} Unix tool 
in order to modify the routing tables of nodes.   
We 
place a jammer such that it affects one or more links. 
Along each route, links that are not affected by the jammer  
consistently use a rate adaptation algorithm. 
{\em On the links that are subject to jamming,
our analysis dictates the decision on whether to use fixed or adaptive rate assignment.} 
We measure the end-to-end throughput on the route. 
We show our results for routes on which, in the absence of a jammer, end-to-end 
throughputs of 6 and 12 Mbps were observed. 
From Figure \ref{fig:route} we see that 
{\em the behavior with rate adaptation on multi-hop routes, 
in the presence of a random jammer, is the same 
as that on a single-hop link}. 
In particular, with low data rates, a sufficiently high PDR has to be sustained over the route,  
in order for a fixed rate approach to perform better than rate adaptation. 
On the other hand, when routes support high data rates, 
fixing the rate on the individual links (that are affected by the jammer) as per our analytical framework, 
provides higher benefits. 

{\bf Choosing the right policy in practice:} To summarize our findings, our analysis demonstrates that using a fixed rate may be attractive on lossless
links while it would be better to use rate adaptation on lossy links. However, as discussed, determining when to use one over the other in real time 
during system operations is difficult;
the determination requires the knowledge of $x(F, R_s)$, $y(R_i)$ and estimates of how often the jammer is active/asleep, on average.
Thus, we choose a simpler practical approach that we call MRC for Markovian Rate Control.
We will describe MRC in detail later (in section \ref{sec:system}) but
in a nutshell, MRC induces memory into the system and keeps track of the feasible rates during benign jamming-free periods; as soon as 
the jammer goes to sleep, legitimate transmissions are invoked at the most recent rate used during the previous sleeping cycle of the jammer.  
We also perform offline measurements by directly
using our analytical formulation (with knowledge of the aforementioned parameters); 
these measurements serve as benchmarks for evaluating the efficacy of MRC (discussed in section \ref{sec:mimo}).

\subsection{Performance of Power Control in the Presence of Random Jamming}
\label{sec:power}
\setcounter{paragraph}{0}

Next, we examine whether tuning power levels can help cope with the interference injected by a jammer. 
If we consider a single legitimate data link and a jammer, incrementing the transmission power on the data link 
should increase the SINR (signal-to-interference plus noise ratio) of the received data packets. 
Thus, one could argue that increasing the transmission power is always beneficial in jamming environments \cite{Lin03}.

We vary the transmission powers of both the jammer and legitimate transceiver, as well as the CCA threshold of the latter. 
Note that the jammer's transmission distribution is not very relevant in this part of our study. 
Our expectation is that tuning the power of legitimate transceivers will provide benefits while the jammer is active. 
{\bf \em In other words, one can expect that the benefits from power control will be similar with any type of jammer.} 
%
%
We define the following:
\begin{itemize}
{
\item
{\boldmath $RSSI_{TR}$} : The RSSI of the signal of the legitimate transmitter at its receiver. 
\item {\boldmath $RSSI_{RT}$}: The RSSI of the signal in the reverse direction (the receiver is now the transmitter). 
\item {\boldmath $RSSI_{JT}$} and {\boldmath $RSSI_{JR}$}:  The RSSI values of the jamming signal at the legitimate transmitter and receiver, respectively. 
\item {\boldmath $RSSI_J$}:  The minimum of  $\{RSSI_{JT}, ~ RSSI_{JR}\}$. 
\item {\boldmath $P_L$} and {\boldmath $CCA_L$}:  The transmission power and the CCA threshold  at legitimate transceivers. 
\item {\boldmath $P_J$}: The transmission power of the jammer.  
}
\end{itemize}
%
%
Our main observations are the following: 
\begin{itemize}
\item 
\textbf
{Mitigating jamming effects by incrementing $P_L$ is viable at low data rates. 
It is extremely difficult to overcome the jamming interference at high rates, \textit{simply with power adaptation}. }
\item 
\textbf{Increasing $CCA_L$ restores (in most cases) the isolated throughput (the throughput achieved in the absence of jammers). }
\end{itemize}
We present our experiments and the interpretations thereof, in what follows.

\subsubsection{Increasing $P_L$ to cope with jamming interference}
\label{sec:simple_power}

Increasing $P_L$ will increase the SINR and one might 
expect that this would reduce the impact of jamming interference on the throughput. 
In our experiments we quantify the gains from employing such a ``brute-force" approach. 

\textbf{\em Details on the experimental process:} 
We perform measurements on 80 different links and with 4 jammers. 
We consider different  fixed values for $P_J$ (from 1 dBm to 18 dBm). 
For each of these values we vary $P_L$ between 1 and 18 dBm 
and observe the throughput in the presence of the jammer, for all possible fixed transmission rates.  
For each chosen pair of values $\{ P_L, P_J \}$,  we run 60-minute repeated experiments and collect a new throughput measurement once every 0.5 seconds. 
Both end-nodes of a legitimate link use the same transmission power. 

\begin{figure*}[t]
\begin{center}
\parbox{2in} {
     \centerline{
\includegraphics[width=4.4cm]{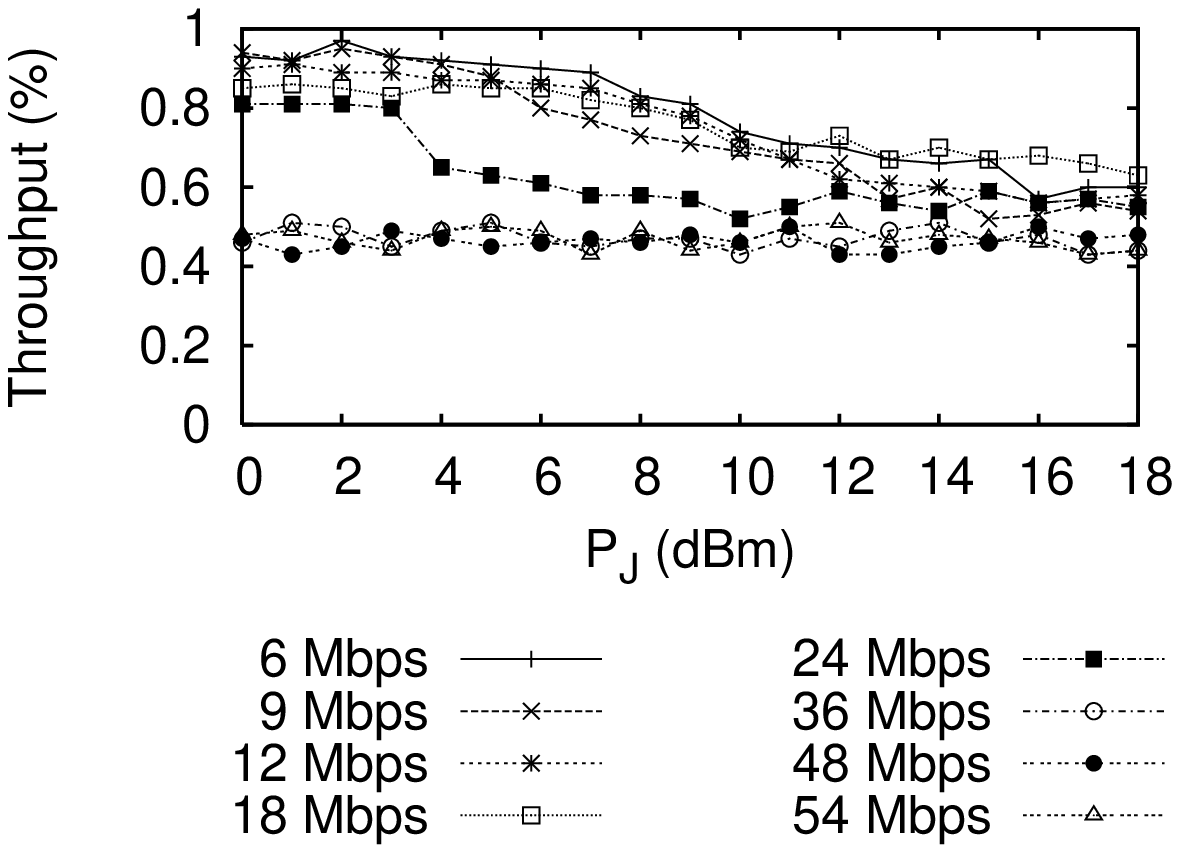}}
\vspace{-0.12in}
     \caption{Percentage of the isolated throughput in the presence of a balanced jammer for various $P_J$ and $P_J$ values and  data rates.}\label{fig:power1}
}
\makebox[.32in] {}
\parbox{2in} {
     \centerline{
\includegraphics[width=5.2cm]{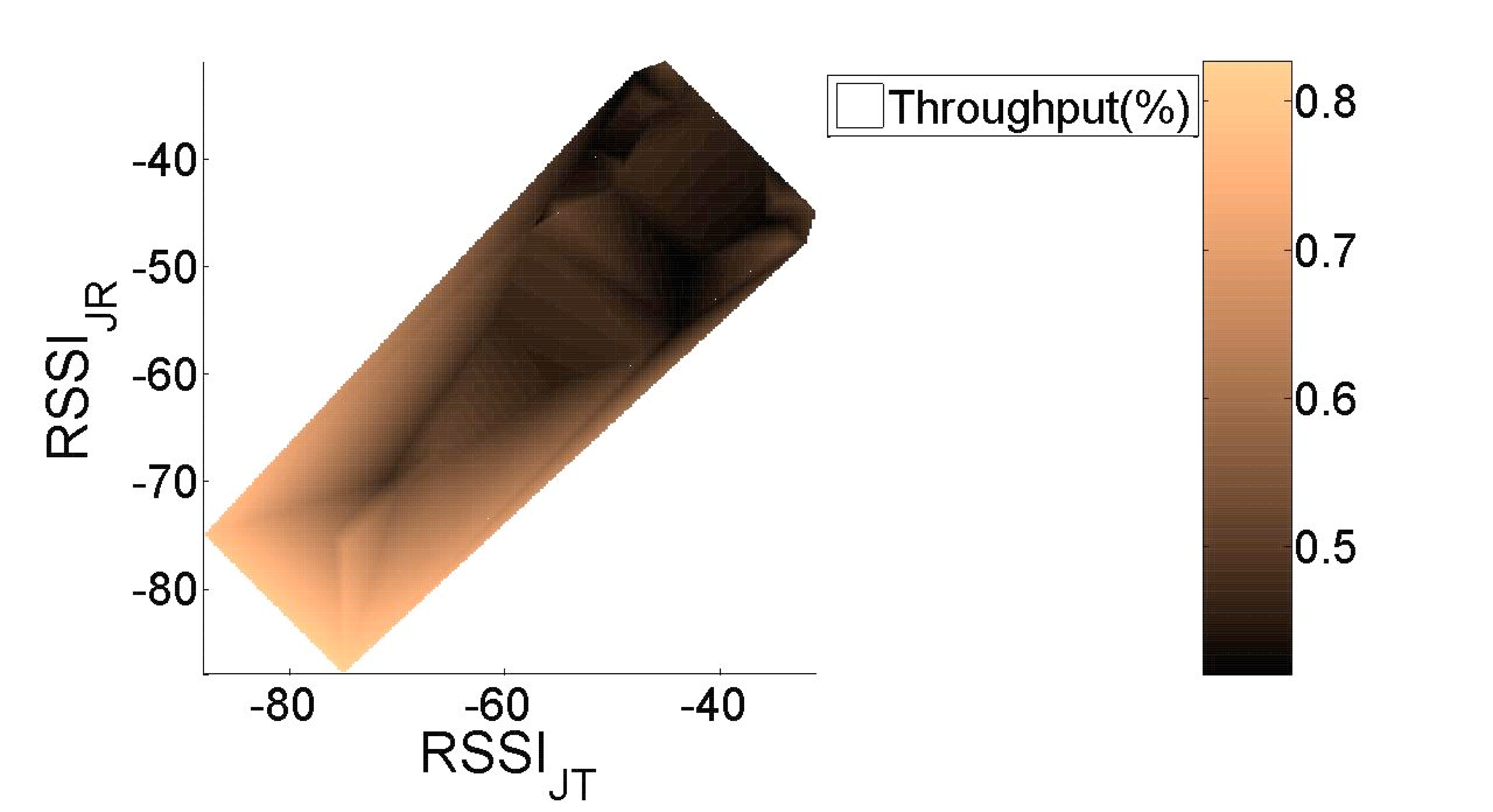}}
\vspace{0.02in}
     \caption{Percentage of the isolated throughput   in the presence of a balanced jammer  Vs. $RSSI_J$,  for $CCA_L$= --80 dBm.}\label{fig:rssi1}
}
\makebox[.32in] {}
\parbox{2in} {
     \centerline{
\includegraphics[width=5.2cm]{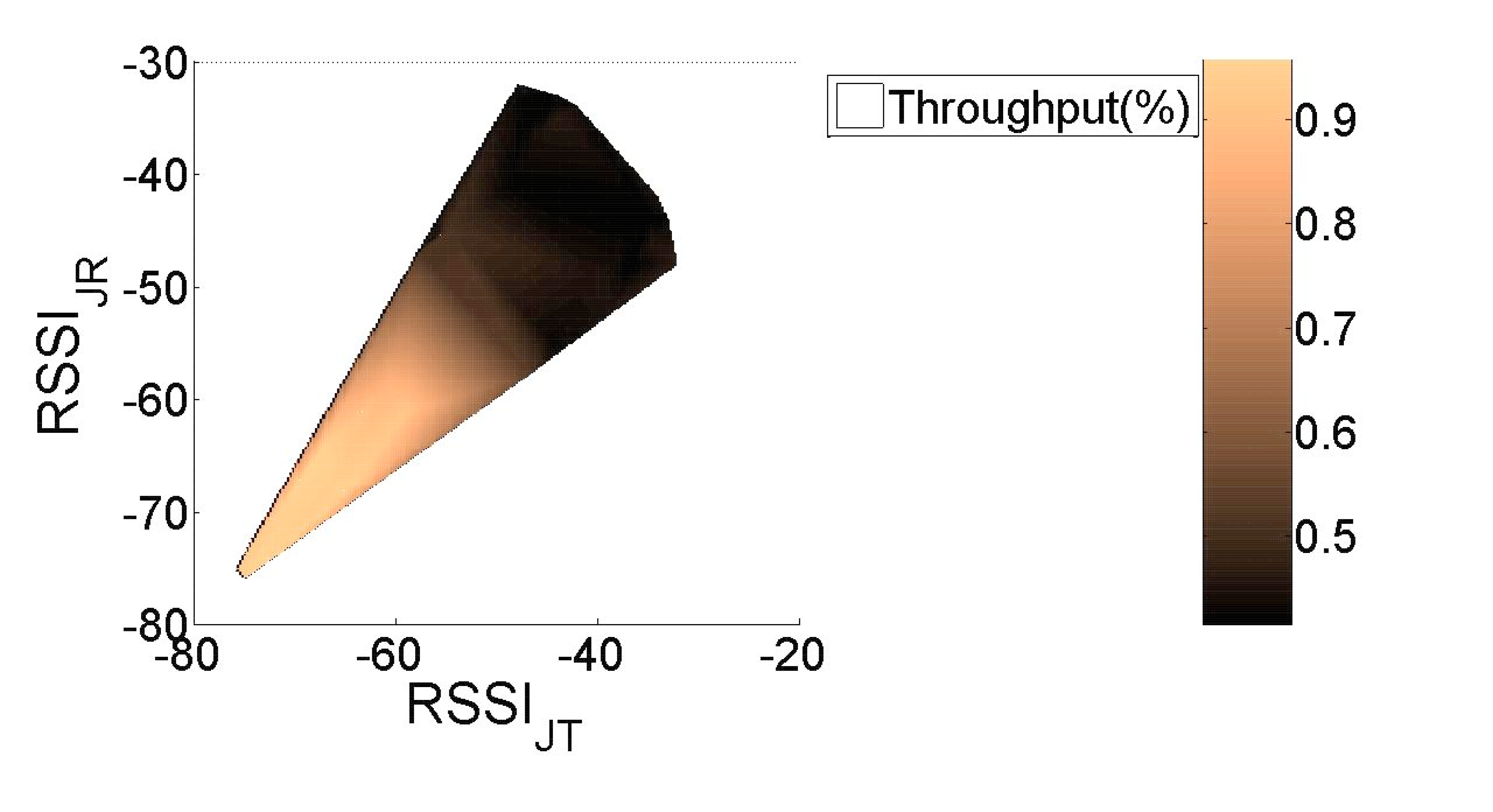}}
\vspace{0.02in}
     \caption{Percentage of the isolated throughput, for various $RSSI_J$ values, and for $CCA_L$ = --50 dBm.}\label{fig:rssi3_50}
}
\end{center}
\end{figure*}

\begin{figure*}[t]
\begin{center}
\parbox{2in} {
     \centerline{
\includegraphics[width=4.4cm]{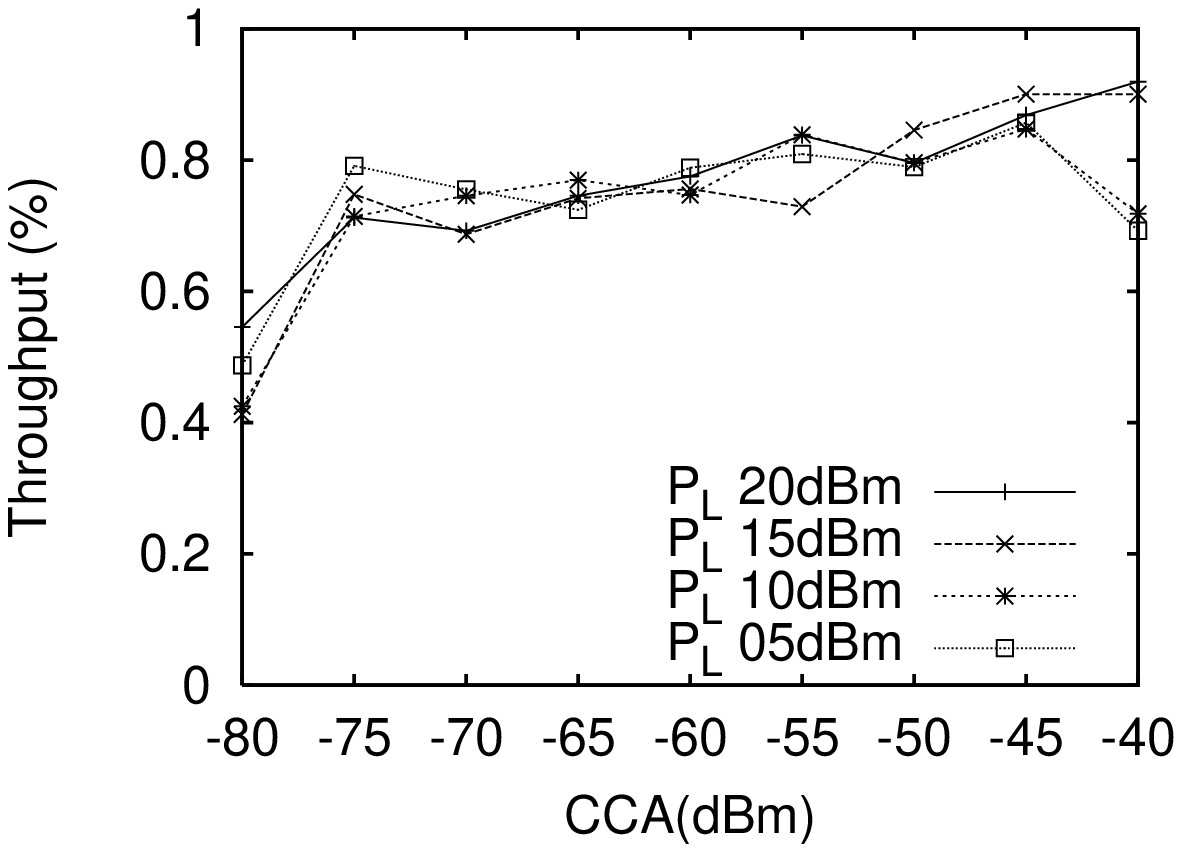}}
\vspace{-0.08in}
     \caption{Percentage of the isolated throughput, for various $CCA_L$ values and various $P_L$ values. $P_J$ = 20  dBm.}\label{fig:cca2}
}
\makebox[.32in] {}
\parbox{2in} {
     \centerline{
\includegraphics[width=3.4cm,angle=270]{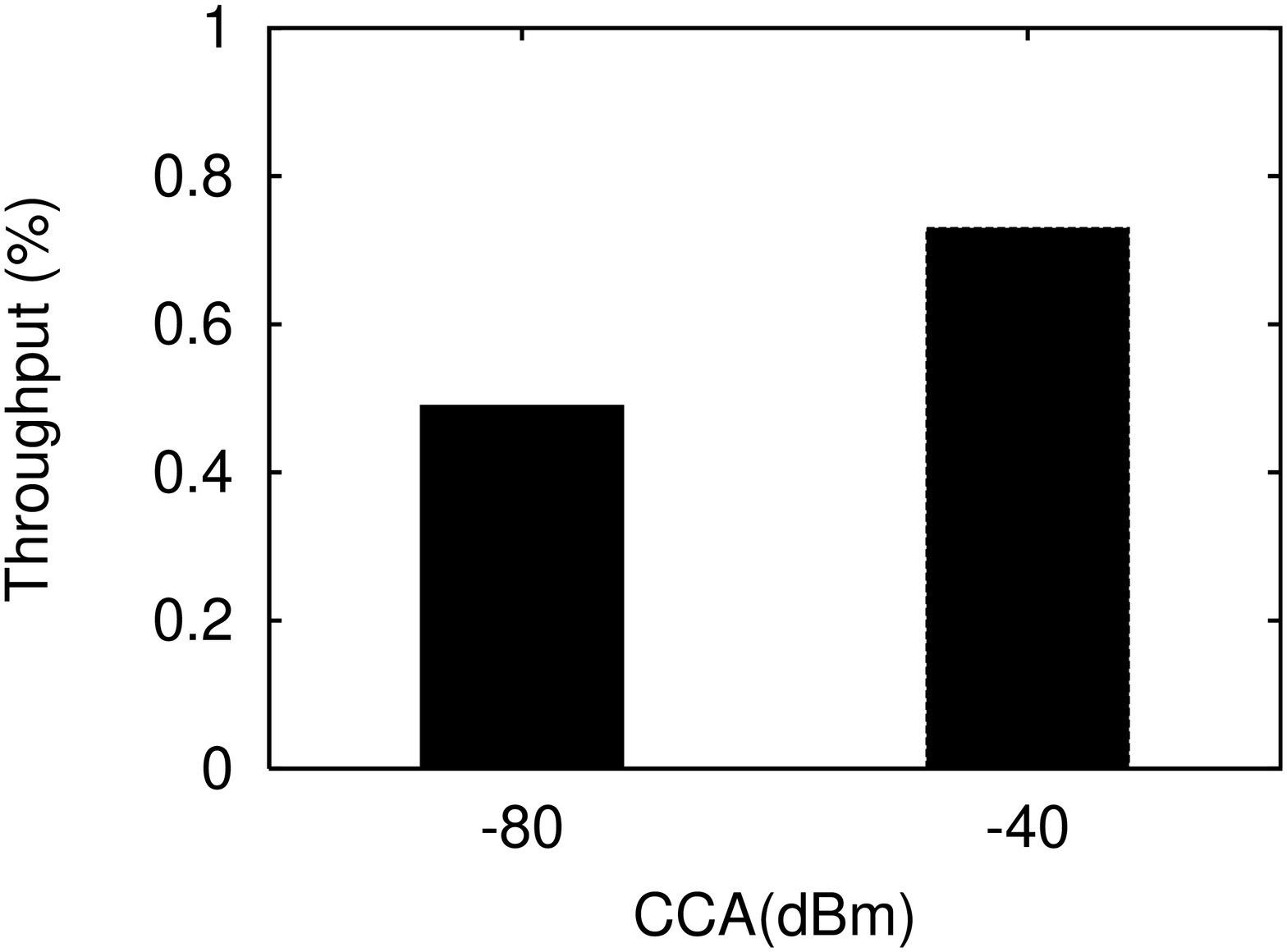}}
\vspace{-0.08in}
     \caption{Careful CCA adaptation significantly improves the end-to-end throughput along a route.}\label{fig:route_cca}
}
\makebox[.32in] {}
\parbox{2in} {\vspace{-0.1in}
     \centerline{
\includegraphics[width=3.4cm,angle=270]{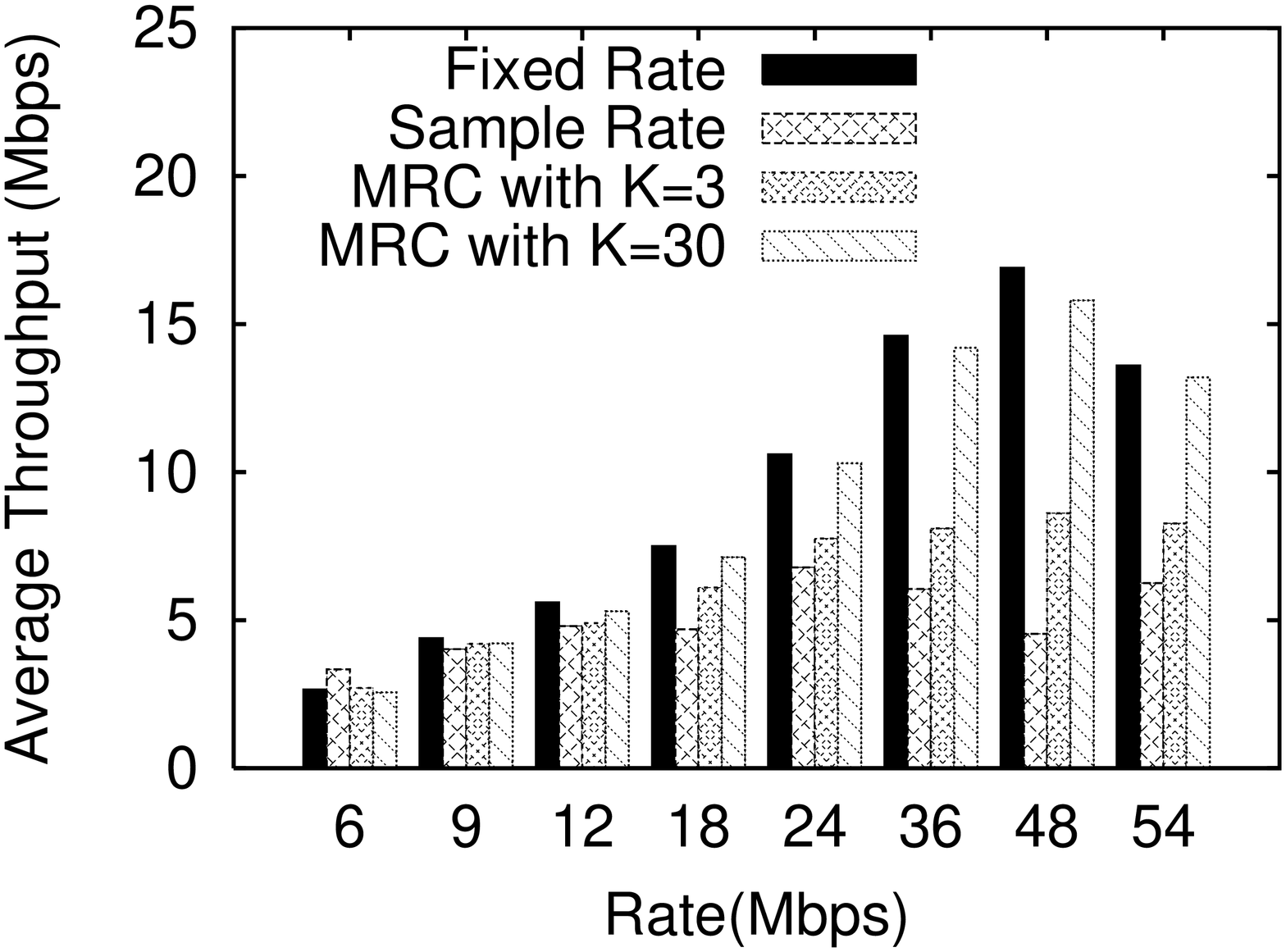}}
\vspace{-0.08in}
     \caption{MRC outperforms current rate adaptation algorithms, especially for high values of $K$.}\label{fig:mrc}
}
\end{center}
\end{figure*}

{\bf The combination of high $P_L$ and low data rate helps mitigate the impact of low-power jammers.} 
We experiment with many different locations of the jammers. 
Our measurements indicate that when high transmission rates are used, increasing $P_L$ does not help alleviate the impact of jammers. 
Sample results are depicted in Figure \ref{fig:rssi_bars}.
In this figure, we plot the percentage of the isolated throughput 
achieved in the presence of jamming, for two representative combinations of $P_L$ and $P_J$ and for 2 different rates.  
In our experiments on the 80 considered links, 
{\em there were no links where incrementing $P_L$ increased the throughput at high data 
rates, 
even with very low jamming powers.} 
While there could exist cases where incrementing $P_L$ could yield benefits at high rates,  
this was not observed.  
In contrast, we observe that with low data rates and when $P_J$ is low, data links can overcome jamming to a large extent by increasing $P_L$.  
Figure \ref{fig:power1} depicts another representative subset of our measurement results  where all legitimate nodes use $P_L$=18 dBm, while $P_J$ is varied between 1 and 18 dBm. 
We observe that the combination of high $P_L$ with low data rate helps overcome the impact of jamming, when $P_J$ is low. 
Note also that when $P_J$ is high, it is extremely difficult to achieve high average throughput. 

The above observations can be explained by taking a careful look at the following two cases: 

{\bf \em  Strong jammer}:  
Let us consider a jammer such that  $RSSI_J$  $> CCA_L$. 
This can result in two effects: 
\textbf{(a)} The sender will sense that the medium is constantly busy and will defer its packet transmissions for prolonged periods of time. 
\textbf{(b)} The signals of both the sender and the jammer will arrive at the receiver with RSSI values higher than $CCA_L$. 
This will result in a packet collision at the receiver.  
In both cases, the throughput is degraded. 
Our measurements show that {\em it is not possible to mitigate strong jammers simply by increasing $P_L$}. 
%

{\bf \em Weak jammer}: 
Let us suppose that the jammer's signals arrive with low RSSI at legitimate nodes.  
This may be either due to energy-conservation strategies implemented by  the jammer 
causing it to use low $P_J$ (e.g., 2 dBm), or due to 
poor channel conditions between a jammer and a legitimate transceiver. 
At high transmission rates, the SINR required for the successful decoding of a packet is larger 
than what is required at low rates (shown in Table \ref{tab:sinrs}) \cite{powInfocom}.  
Our throughput measurements show that even in the presence of  weak jammers,  the 
SINR requirements at  high transmission rates are typically not satisfied. 
However, since the SINR requirements at lower data rates are less stringent, 
{\em the combination of high $P_L$ and low rate, provides significant throughput benefits}. 
\begin{table}[h]
\small{
\centering
\begin{tabular}{|c|c|c|c|c|c|c|c|c|c| l|r|}
\hline
Data Rate & 6 & 9 & 12 & 18 & 24 & 36 & 48 & 54\\	
\hline
SINR (dB) & 6 & 7.8 & 9 & 10.8 & 17 & 18.8 & 24 & 24.6\\
\hline
\end{tabular} 
\caption{SINR levels required for successful packet decoding, in 802.11a/g.}
\label{tab:sinrs}
}
\end{table}

\subsubsection{Tuning  $CCA_L$ on single-hop settings}
\label{sec:cca}

Next, we investigate the potential of adjusting $CCA_L$ in conjunction with $P_L$. 
 
\textbf{\em Implementation and experimental details:} 
For these experiments we exclusively use the {\em Intel-2915} cards; these cards allow us to tune the CCA threshold. 
We have modified a prototype version of the AP/client driver, in order to periodically collect measurements for $RSSI_{TR}$, $RSSI_{RT}$ and $RSSI_J$.  
We consider $80$ AP-client data links, with traffic flowing from the AP to the client. 
As before, we divide the 80 data links into 20 sets of 4 isolated links. 
We use Intel's proprietary rate adaptation algorithm, which has been implemented in the firmware of the {\em Intel-2915} cards. 
We measure the achieved data throughput for different values of $P_L$ and $CCA_L$. 
Both nodes of a data link use the same power and CCA threshold values.

{\bf Tuning the CCA threshold is a potential jamming mitigation technique.} 
To begin with, we perform throughput measurements with the default $CCA_L$ value (-80 dBm), and with 
various $RSSI_J$ values.  
We observe from Figure \ref{fig:rssi1} that when 
$RSSI_J < CCA_L$,
data links achieve high throughputs.  
This is because signals with RSSI $< CCA_L$ are ignored by the transceiver's hardware. 
In particular, 
	{\bf (a)} such signals do not render the medium busy, and 
	{\bf (b)} receivers are trying to latch onto 
signals with RSSI $> CCA_L$, while other signals are considered to be background noise.  
Moreover, even when $RSSI_J$ is slightly larger than $CCA_L$, we still observe decent throughput achievements for the cases wherein data links operate at high SINR regimes. 
These measurements imply that the ability to tune $CCA_L$ can help receive data packets correctly, even while jammers are active. 


In order to further explore the potential of such an approach, 
we vary $CCA_L$ from -75 to -30 dBm  on each of the considered 80 links. 
Figure 
 \ref{fig:rssi3_50} depicts the results for the case where  $CCA_L$ is  equal to 
-50 dBm.
 We observe that {\bf \em increasing $CCA_L$ results in significantly higher data throughputs,  
even with quite high $RSSI_J$ values.} 
More specifically, from Figure 
\ref{fig:rssi3_50} we observe that when $RSSI_J$ 
is lower than $CCA_L$, links can achieve up to 95\% of the throughput that is achieved when the link operates in isolation (jamming-free).  
When $RSSI_J \approx CCA_L$, data links still achieve up to 70\% of the jamming-free throughput (capture of data packets is still possible to a significant extent).  
As one might expect, if  $RSSI_J \gg CCA_L$,  there are no performance benefits. 


Our observations  also hold in some scenarios where, $P_J > P_L$. 
Figure  \ref{fig:cca2} presents the results from one such scenario. 
We observe that appropriate CCA settings can allow legitimate nodes to exchange traffic effectively, 
even when $P_J \gg P_L$. 
This is possible if the link conditions between the jammer and the legitimate transceivers   
are poor and result in low $RSSI_J$.  
Note here that one cannot increase $CCA_L$ to arbitrarily high values on legitimate nodes. 
Doing so is likely to compromise connectivity between nodes or degrade the throughput due to failure of capturing packets as seen in Figure \ref{fig:cca2} for $P_L=5dBm$ and $P_L=10dBm$.

\subsubsection{Tuning $CCA_L$ in multi-hop configurations}
\label{sec:ccamul} 
We perform experiments with various CCA thresholds along a route. 
Previous studies have shown that in order to avoid  starvation 
due to 
asymmetric links, 
the transmission power and the CCA threshold need to be jointly tuned for all nodes of the same connected (sub)network \cite{powInfocom}. 
In particular, the product $C = P_L \cdot CCA_L$ must be the same for all nodes. 
Given this, we ensure that $C$ is the same for all nodes that are part of  a route. 
In particular, we set $P_L$ to be equal to the maximum possible value of  20 dBm on all nodes of a route;  
for each run, $CCA_L$ is therefore set to be the same on all of the nodes on the route. 
Throughout our experiments with multi-hop traffic, nodes on  one route do not interfere with nodes that are on other routes. 
In scenarios where nodes belonging to different routes interfere with each other,  if all nodes use the same $P_L$,  
their $CCA_L$ values must be the same \cite{powInfocom}, \cite{mdg}. 
However, we did not experiment with such scenarios given that our objective is to isolate the impact of a jammer and not to examine interference between coexisting sessions in a network. 

We experiment with the same multi-hop settings as in section \ref{sec:multihoprate}. 
Figure \ref{fig:route_cca} presents the results observed on one of our routes. 
We observe that careful CCA tuning can provide significant average end-to-end throughput benefits along a route. 

\section{Designing ARES} 
\label{sec:system}
\setcounter{paragraph}{0}

In this section, we design our system ARES based on the observations from the previous section. 
ARES is composed of two main modules:
	{\bf (a)} a {\em rate module} that decides between fixed or adaptive-rate assignment, and  
	{\bf (b)} a {\em power control module} that facilitates appropriate CCA tuning on legitimate nodes. 

{\bf Rate Module in ARES:}
As discussed in section \ref{sec:ratemeas}, our experiments with three popular rate adaptation algorithms show that 
the convergence time of the algorithms affects the link performance in random-jamming environments. 
This convergence time is largely implementation specific. 
As an example, our experiments with both SampleRate and Onoe show that in many cases it takes more than 10 sec for both  algorithms to converge to the ``best" rate; \cite{bug} reports similar observations. 
The rate module in ARES decides on whether a fixed or an adaptive-rate approach should be applied. 

\textbf{\em MRC: Markovian Rate Control:} 
MRC is an algorithm--patch that can be implemented on top of any rate control algorithm. 
MRC is motivated by our analysis in section \ref{sec:rate}. 
However, as discussed earlier, it does not directly apply the analysis, since this would require extensive offline measurements (the collection of which can be time-consuming)
and estimates of the jammer active and sleep periods.  
The key idea that drives MRC is that
 a rate adaptation algorithm need not examine the performance at all the transmission rates during the sleeping period of the jammer. 
The algorithm simply needs to remember the previously used transmission rate, and use it as soon as the jammer goes to sleep. 
Simply put, MRC introduces {\em memory} into the system. The system keeps track of past transmission rates and hops to the 
stored 
highest-rate state 
as soon as 
the jammer goes to sleep. 
Since the channel conditions may also change due to the variability in the environment, MRC invokes the re-scanning of all rates periodically, 
once every $K$ consecutive sleeping/jamming cycles. 
When $K=1$ we do not expect to have any benefits, since the scanning takes place in each cycle. 

Note here that the appropriate value of $K$ depends on the environment and the sleep and active periods of the jammer. One could adaptively
tune the $K$ value. As an example, an additive increase additive decrease strategy may be used
where one would increase the value of $K$ until a degradation is seen.
The $K$ value would then be decreased. The implementation of such a strategy is beyond the scope of this paper and will be considered in the future.

{\em Implementation details of MRC:} 
The implementation 
	{\bf (a)} keeps track of the highest transmission rate used over a benign time period (when the jammer is asleep) and, 
	{\bf (b)} applies this rate immediately upon the detection of the next transition from the jammer's active period to the sleeping period. 

Figure \ref{fig:mrc} presents a set of measurements with MRC, with intermittent SampleRate invocations (once every $K$  cycles) for $K = \{3, 30\}$.  
We observe that MRC outperforms pure SampleRate in jamming environments, especially with larger values of $K$. 
With small $K$, the rate adaptation algorithm 
is invoked often and this reduces the achieved benefits.  
Furthermore, 
MRC provides throughput that is 
close to the maximum achievable 
on the link (which may be either with fixed or adaptive rate, depending on whether the link is lossy or lossless).  



%
\begin{figure*}[ht]
\begin{center}
\includegraphics[width=14cm]{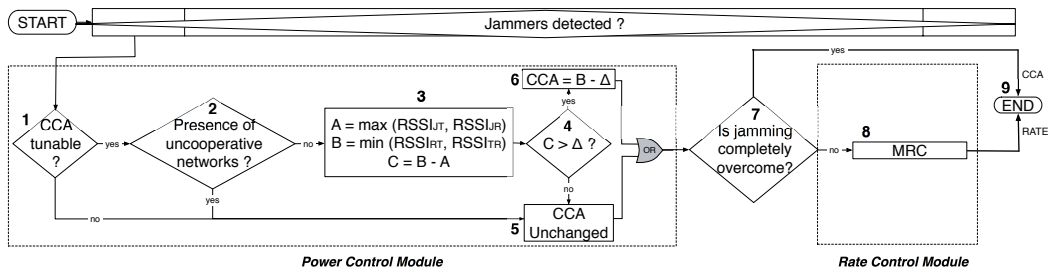} \vspace{-0.1in}
\vspace{-0.1in}
\caption{ARES: our Anti-jamming Reinforcement System.} \label{fig:ares}
\end{center}
\end{figure*} 

{\bf Power Control Module in ARES:}
As discussed in section \ref{sec:power}, increasing $P_L$ is beneficial at low rates; 
while at high rates this is not particularly useful, it does not hurt either.
Since our goal in this paper is to propose methods for overcoming the effects of  jamming (and not legitimate) interference, we 
impose the use of the maximum $P_L$  by all nodes in the presence of jammers. 
The design of a power control mechanism that in addition takes into account the imposed legitimate interference (due to high $P_L$) is beyond the scope of this paper. 

More significantly, our power control module overcomes jamming interference by adaptively tuning $CCA_L$. 
The module requires the following inputs on each link: 
\begin{itemize}
\item The values of $RSSI_{TR}$, $RSSI_{RT}$, $RSSI_{JR}$, and $RSSI_{JT}$.  
These values can be easily observed in real time.
%
%
\item An estimation for the shadow fading variation of the channel, $\Delta$. 
Due to shadow fading, the above RSSI values can occasionally vary by $\Delta$. 
The value of $\Delta$ is dependent on the environment of deployment. 
One can perform offline measurements and configure the value of $\Delta$ in ARES.
\end{itemize}
We determine the variations in RSSI measurements via experiments on a large set of links. 
The measurements indicate that $\Delta$ is approximately 5 dB for our testbed (a less conservative value than what is reported in \cite{zvan}). 
The value of $CCA_L$ has to be at least $\Delta$ dB lower than both $RSSI_{TR}$ and $RSSI_{RT}$, to guarantee connectivity at all times. 
Hence, ARES sets: 
$$CCA_L = min(RSSI_{TR}, RSSI_{RT}) - \Delta, ~~~~ \text{if}$$ 
$$max(\hspace{-0.5mm} RSSI_{JT},\hspace{-0.5mm} RSSI_{JR}\hspace{-0.5mm})\hspace{-0.5mm} \leq \hspace{-0.5mm} min(\hspace{-0.5mm}RSSI_{TR},\hspace{-0.5mm} RSSI_{RT}\hspace{-0.5mm}) - \Delta.$$ 
Otherwise, $CCA_L$ is not 
changed\footnote{We choose not to tune $CCA_L$, unless we are certain that it can help alleviate jamming interference.}. 
This ensures that legitimate nodes are always connected, while the jammer's signal is ignored to the extent possible.  
Our experiments indicate that, especially if 
$$max(\hspace{-0.5mm}RSSI_{JT},\hspace{-0.5mm} RSSI_{JR}\hspace{-0.5mm})\hspace{-0.5mm} \leq \hspace{-0.5mm}min(\hspace{-0.5mm}RSSI_{TR},\hspace{-0.5mm} RSSI_{RT}\hspace{-0.5mm}) - 2\Delta, $$
the data link  can operate as if it is jamming-free.   

In order to avoid starvation effects, the tuning of the CCA threshold should be performed only when 
nodes that participate in power control belong to the same network \cite{mdg}.  
Unless collocated networks cooperate in jointly tuning their CCA (as per our scheme), 
our power control module will not be used.  
Note that when jamming attacks become more prevalent, cooperation between coexisting networks may be essential
in order to fight the attackers.
Hence, in such cases collocated networks can have an agreement to jointly increase the CCA thresholds when there is a jammer.  

\textbf{\em Implementation details:} 
Our power control algorithm can be applied in a {\bf centralized} manner by having all legitimate nodes report 
the required RSSI 
values to a central server. 
The central server then applies the same $CCA_L$ value to all nodes (of the same connected network). 
The chosen $CCA_L$ is the highest possible CCA threshold that guarantees connectivity between 
legitimate nodes. 
This reporting  
requires trivial modifications on the wireless drivers.  
We have implemented a centralized functionality when our network is configured as a multi-hop wireless mesh.

In a {\bf distributed} setting, our algorithm is applicable as long as legitimate nodes are able to exchange RSSI information. 
Each node can then independently determine the $CCA_L$ value. 
To demonstrate its viability, we implement and test a distributed version of the power control module in a 802.11a/g WLAN configuration. 
In particular, we modify the Intel prototype AP driver, 
by adding an extra field in the ``Beacon" template. 
This new field contains a matrix of RSSI values of  neighboring jammers and legitimate nodes. 
We enable the decoding of received beacons in the AP driver (they do not read these by default). 
Assuming that a jammer imposes almost the same amount of interference on all devices (AP and clients) within a cell, the AP of the cell determines 
the final $CCA_L$ after a series of iterations in a manner very similar to the approaches in \cite{mdg}, \cite{powInfocom}. 

{\bf Combining the modules to form ARES:}
We combine our rate and power control modules to construct ARES 
as shown in Figure \ref{fig:ares}. 
ARES also includes a jamming detection functionality. Towards this we incorporate a mechanism that was proposed in \cite{Xu05};  
this functionality  performs a consistency check between the instantaneous PDR and RSSI values. 
If the PDR is extremely low while the RSSI is much higher than the default $CCA_L$, the node is considered to be jammed. 

The goal of ARES is to detect jammers and apply the individual modules as appropriate. 
ARES applies the power control module first, since with this module, the impact of the jammer(s) could be completely overcome. 
If the receiver is able to capture and decode all packets in spite of the jammer's transmissions, no further actions are required. 
Note that even if $CCA_L > RSSI_J$, the jammer can still affect the link performance. 
This is because 
with CCA tuning  
 the jamming signal's power is added to the noise power.  
Hence, 
even though the throughput may increase, the link may not achieve the ``jamming-free performance" while the jammer is active.  
If the jammer still has an effect on the network performance after tuning $CCA_L$, (or if CCA tuning is infeasible due to the
presence of collocated uncooperative networks) ARES enables the rate module. 
Note that the two modules can operate independently and the system can bypass any of them in case the hardware/software does not support the specific functionality.

\section{Evaluating our system} 
\label{sec:mimo}
\setcounter{paragraph}{0}

We first evaluate ARES by examining its performance 
in three 
different networks:  
a MIMO-based WLAN, 
a mesh network in the presence of mobile-jammers, and
an 802.11a WLAN setting where uplink TCP traffic is considered.  

{\bf ARES boosts the throughput of our MIMO WLAN under jamming by as much as 100\%:}
Our objective here is twofold. First, we seek to observe and understand the behavior 
of MIMO networks in the presence of jamming. Second, we wish to measure the effectiveness of  ARES in such settings. 
Towards this, we deploy a set of 7 nodes equipped with \textit{Ralink RT2860} miniPCI cards. 

\textbf{\em Experimental set-up:} 
We examine the case for a WLAN setting, since the {\em RT2860} driver does not currently support the ad-hoc mode of operations. 
MIMO links with Space-Time Block Codes (STBC) are expected to provide  robustness to signal variations,  
thereby reducing the average SINR that is required for achieving a desired bit error rate, 
as compared to a corresponding SISO (Single-Input Single-output) 
link.
We modify the client driver of the cards to enable 2$\times$2 STBC support.
This involves adding the line  
$$\texttt{\{"HtStbc",~~~ Set\_HtStbc\_Proc\}}$$   
into the {\em RTMP\_PRIVATE\_SUPPORT\_PROC} struct array, located in $\texttt{os/linux/sta\_ioctl.c}$ in the driver. 
We consider 2 APs, with 2 and 3 clients each, and two jammers. 
Fully-saturated downlink UDP traffic flows from each AP to its clients. 

\textbf{\em Applying ARES on a MIMO-based WLAN:} 
We first run experiments without enabling ARES. 
Interestingly, we observe that in spite of the fact that STBC is used, 802.11n links present the same vulnerabilities as 802.11a or g links.  
In other words, MIMO does not offer significant benefits by itself, in the presence of a jammer. 
This is due to the fact that 802.11n is still employing CSMA/CA and as a result the jamming signals can render the medium busy for a MIMO node as well. 
%
%
%
%
Moreover, for STBC codes to work effectively and provide
a reduction in the SINR for a desired bit error rate (BER), 
the signals received on the two antenna elements will have to experience independent multipath fading effects. 
In other words, a line of sight or dominant path must be absent. 
However, in our indoor testbed, given the proximity of the communicating transceiver pair, this may not be the case.  
Thus, little diversity is achieved \cite{jafar} and does not suffice in coping with the jamming effects.

\begin{figure*}[ht]
\begin{center}
\parbox{1.5in} {
     \centerline{
\includegraphics[scale=.18,angle=270]{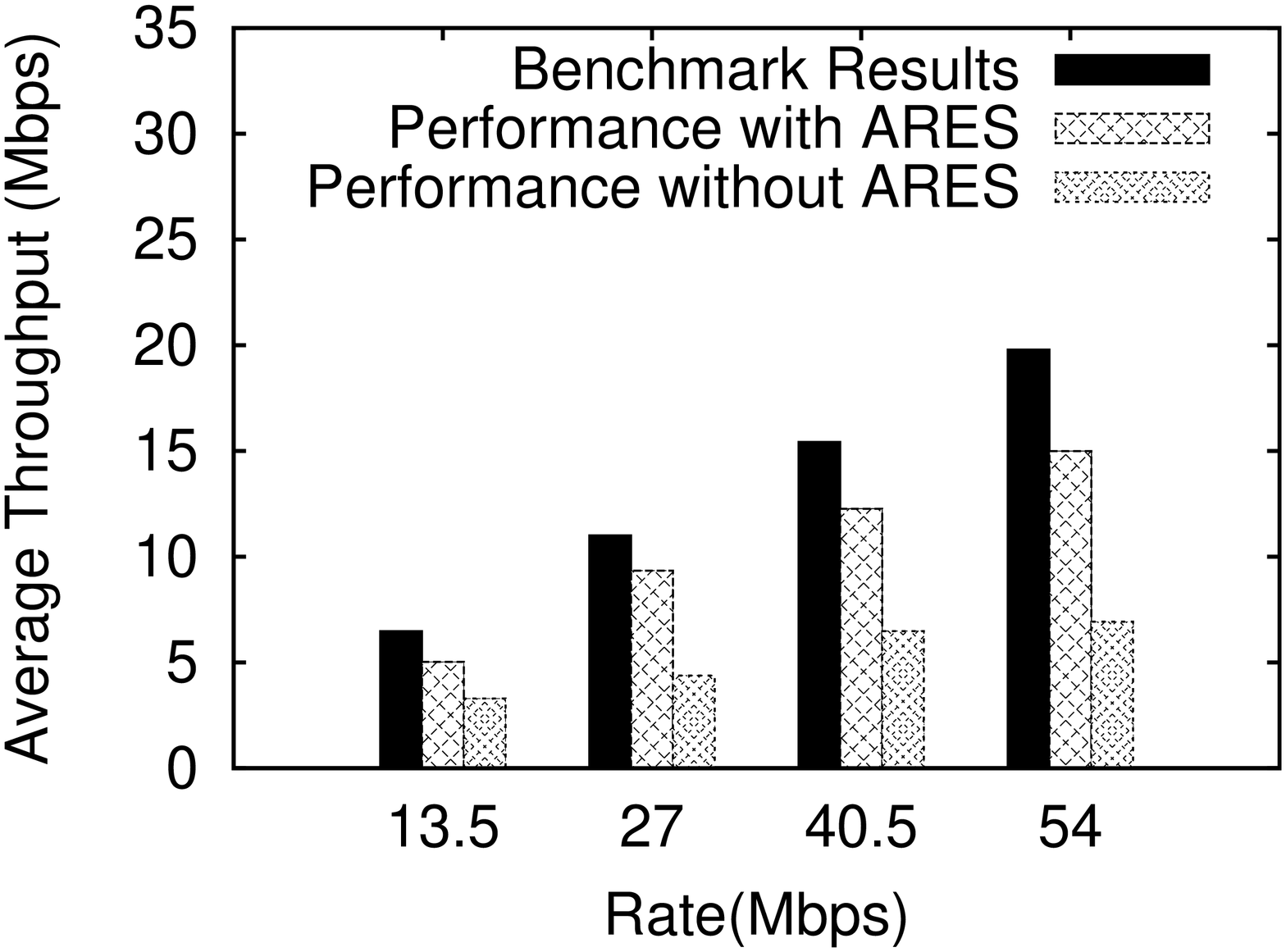}} \vspace{-0.0752in}
     \caption{ARES provides significant throughput benefits in a MIMO network in the presence of jammers.}\label{fig:mimofig}
}
\makebox[.2in] {}
\parbox{1.5in} {
     \centerline{
\includegraphics[scale=.18,angle=270]{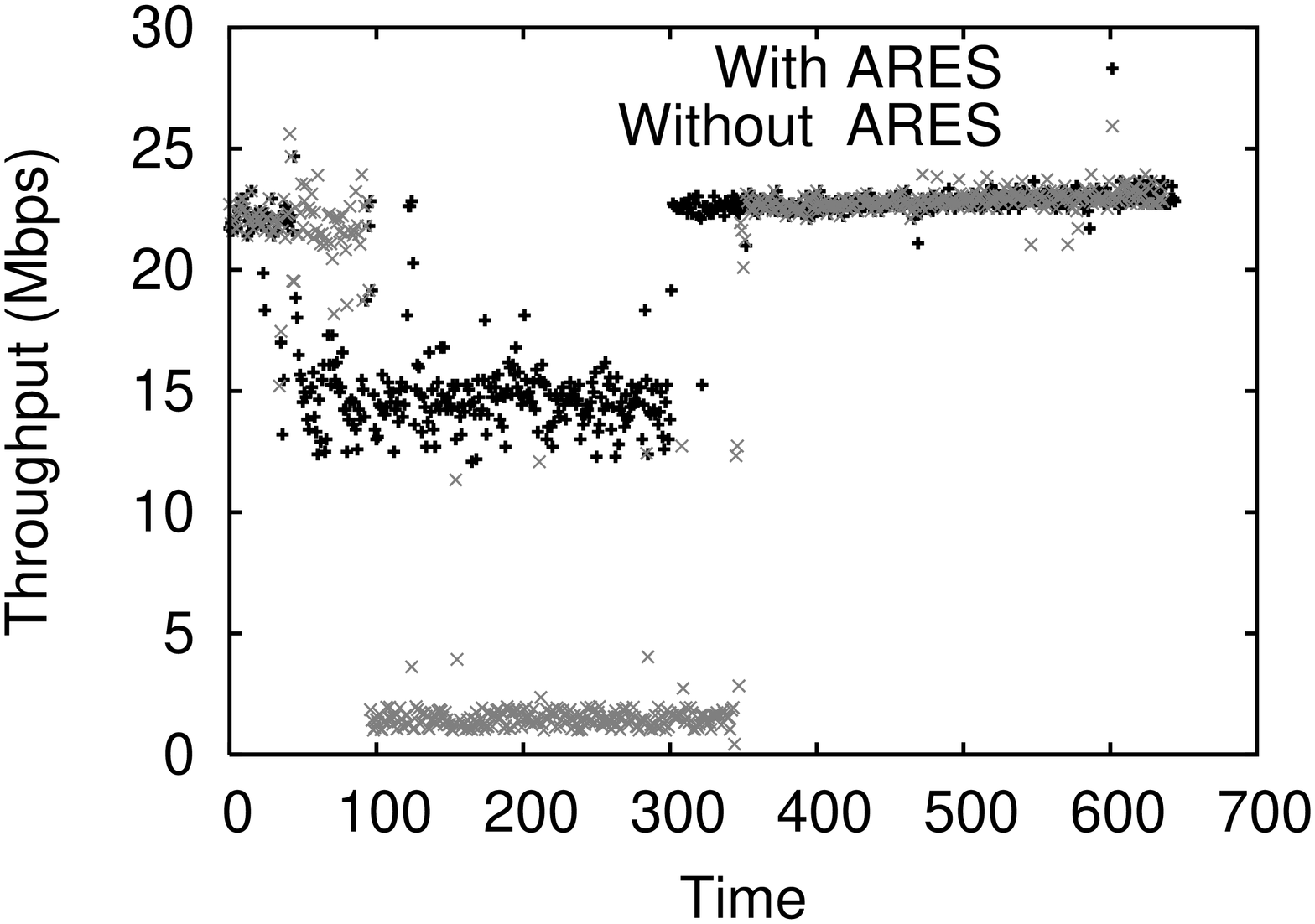}} \vspace{-0.0752in}
     \caption{ARES provides significant throughput improvement in mobile-jamming scenarios.}\label{fig:mobijam}
}
\makebox[.2in] {}
\parbox{1.5in} {
     \centerline{\hspace{-0.4cm} 
\includegraphics[scale=.18,angle=270]{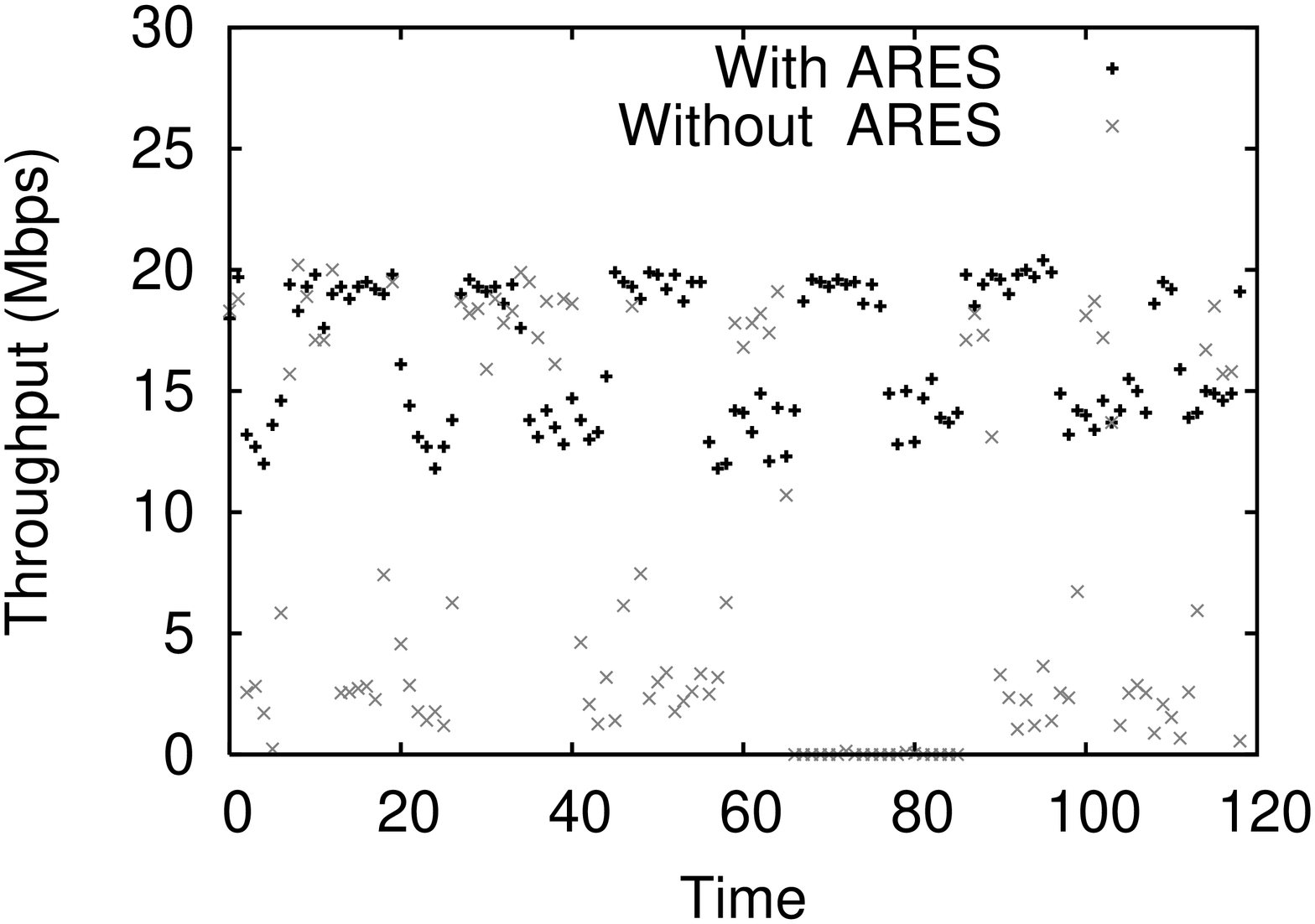}} \vspace{-0.0752in}
     \caption{ARES improves the client-AP link throughput by 130\% with TCP traffic scenarios.}\label{fig:wlantcp}
}
\makebox[.2in] {}
\parbox{1.5in} {
     \centerline{\hspace{-0.4cm} 
\includegraphics[scale=.18,angle=270]{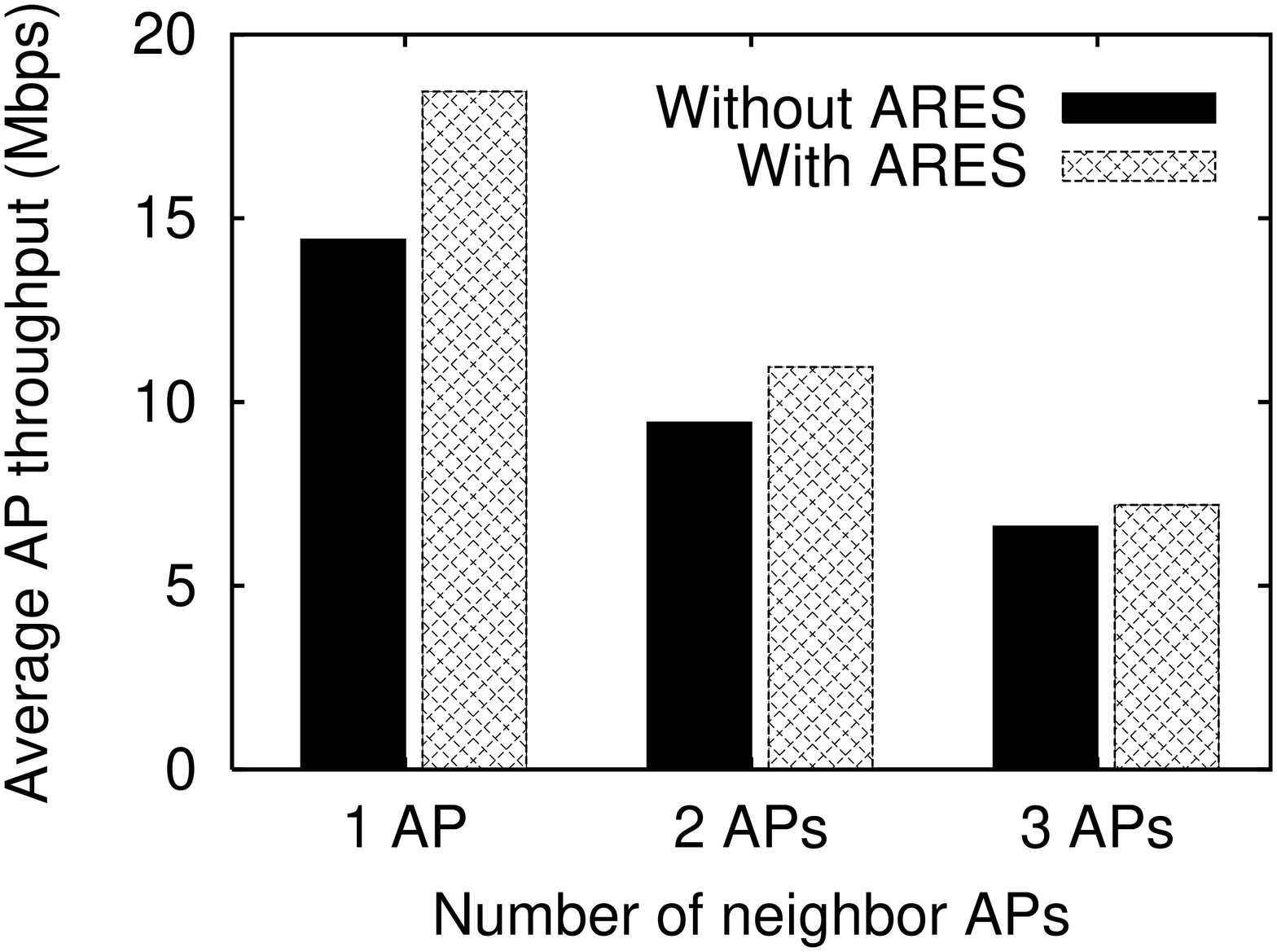}} \vspace{-0.0752in}
     \caption{MRC improves the throughput of neighbor legitimate devices, as compared to SampleRate.}\label{fig:mrcNET}
}
\end{center}
\end{figure*}

Next, we apply ARES and observe the behavior. 
The path that ARES follows (in Figure \ref{fig:ares}) is 
$1 \rightarrow 5 \rightarrow 7 \rightarrow 8 \rightarrow 9$. 
Since the CCA threshold is not tunable with the {\em RT2860} cards, ARES derives decisions with regards to rate control only. 
Figure \ref{fig:mimofig} depicts the results.  
We observe that the configuration with ARES 
outperforms the rate adaptation scheme that is implemented on the {\em RT2860} cards in the presence
of the jammer, by as much as 100\%. 
Note that higher gains would be possible, if ARES was able to invoke the  power control module. 

In Figure \ref{fig:mimofig}
we also compare the throughput with MRC against the suggested settings with our analysis (these settings allow us to obtain benchmark measurements
possible with global information).  
The parameters input to the analysis 
are the following: 
{\bf (a)} The jammer is balanced with a jamming distribution $U[1, 5]$ and a sleep distribution $U[1, 6]$. 
{\bf (b)} We examine 4 $R_a$ values: 13.5, 27, 40.5, 54 Mbps. 
{\bf (c)} $F$ = 0 Mbps.  
{\bf (d)} We input estimates of the $y(R_i)$ values which are obtained via comprehensive offline measurements. 
{\bf (e)} The offline measured $PDR_f$. 
We observe that the performance with MRC 
is quite close to our benchmark measurements. These results show that in spite of having no information with regards to the jammer distribution or
the convergence times of the rate adaptation algorithms, MRC is able to significantly help in the presence of a random jammer.


{\bf ARES increases the link throughput by up to 150\% in a mesh deployment with mobile jammers:}
Next, we apply ARES in an 802.11a/g mesh network with mobile random jammers.  
We also consider a frequent jammer (jamming distribution $U[1, 20]$ and sleeping distribution $U[0, 1]$). The jammer moves towards the 
vicinity of the legitimate nodes, 
remains there for $k$ seconds, and subsequently moves away. 
For the mobile jammer we used a laptop, equipped with one of our Intel cards, and  carried it around.
The power control module is implemented in a centralized manner. 
ARES increases $CCA_L$ in order to overcome the effects of jamming interference, to the extent possible. 
In this case, due to the aggressiveness of the considered jammer (prolonged jamming duration), the rate adaptation module does not provide any benefits (since rate control helps only when the jammer is sleeping). 
In this scenario, ARES follows the path: 
$1 \rightarrow 2 \rightarrow 3 \rightarrow 4 \rightarrow 6 \rightarrow 7 \rightarrow 8 \rightarrow 9$.
Figure \ref{fig:mobijam} depicts throughput-time traces, with and without ARES, for  an arbitrarily chosen link and $k \approx 200$. 
The use of ARES tremendously increases the link throughput during the jamming period (by as much as 150 \%). 
We have observed the same behavior with a distributed implementation of the power control module in an 802.11a WLAN setting. 

{\bf ARES improves the total AP throughput by up to 130\% with TCP traffic:} 
Next, we apply ARES on a 802.11a WLAN. 
For this experiment, we use nodes equipped with the {\em Intel-2915} cards. 
We consider a setting with 1 AP and 2 clients, where clients can sense each others' transmissions. 
We place a balanced jammer (jamming distribution $U[1, 5]$ and sleeping $U[1, 8]$) such that 
all 3 legitimate nodes can sense its presence. 
We enable fully-saturated uplink TCP traffic from all clients to the AP (using {\em iperf}) and we measure the total throughput at the AP, once every 0.5 sec. 
In this scenario, ARES follows the path: 
$1 \rightarrow 2 \rightarrow 3 \rightarrow 4 \rightarrow 6 \rightarrow 7 \rightarrow 8 \rightarrow 9$.
From Figure \ref{fig:wlantcp}, 
we observe that {\em the total AP throughput is improved by up to 130\% during the periods that the jammer is active}. 
The benefits are less apparent when the jammer is sleeping 
because TCP's own congestion control algorithm is unable to fully exploit the advantages offered by the fixed rate strategy.

{\bf Applying MRC on an AP improves the throughput of neighbor APs by as much as 23\%:} 
With MRC, a jammed node utilizes the lowest rate (when the jammer is active) and highest rate (when the jammer is sleeping) that provide the maximum long-term throughput. 
With this, the jammed node avoids examining the intermediate rates and, as we showed above, this increases the link throughput. 
We now examine how this rate adaptation strategy affects the performance of neighbor legitimate nodes. 
We perform experiments on a topology consisting of 4 APs and 8 clients, with 2 clients associated with each AP, all set to 802.11a mode. 
A balanced jammer with a jamming distribution $U[1, 5]$ and a sleep distribution $U[1, 6]$ is placed such that affects {\em only} one of the APs. Only the affected AP is running MRC; the rest of the APs use SampleRate. 
We activate different numbers of APs at a time, and we enable fully-saturated downlink traffic from the APs to their clients. 
Figure \ref{fig:mrcNET} depicts the average total AP throughput. 
Interestingly, we observe that {\em the use of MRC on jammed links improves the performance of neighbor APs that are not even affected by the jammer}. 
This is because the jammed AP does not send any packets using intermediate bit rates (such as with the default operation of rate adaptation algorithms). 
Since MRC avoids the transmission of packets at lower (that the highest sustained) bit rates, the jammed AP does not occupy the medium for as prolonged periods as with the default rate control techniques; the transmission of packets at the high rate (while the jammer is asleep) takes less time. 
Hence, this provides more opportunities for neighbor APs to access the medium, thereby increasing the AP throughput. 
Specifically, we observe that the throughput of one neighbor AP is improved by 23\% (when the topology consists of only 2 APs, one of which is jammed).  
As we further increase the number of neighbor APs, the benefits due to MRC are less pronounced, due to increased contention (Figure \ref{fig:mrcNET}). 
We elaborate of the efficacy of MRC in the following section.

{\bf ARES converges relatively quickly:} 
Finally, we perform experiments to assess how quickly the distributed form of ARES converges to a rate and power control setting. 
In a nutshell, our implementation has demonstrated that the network-wide convergence time of ARES is relatively small.  
With MRC, the rate control module can very rapidly make a decision with regards to the rate setting; as soon as the jammer is detected, MRC applies the appropriate stored lowest and highest rates. 

With regards to the convergence of the power control module, 
recall that our implementation involves the dissemination of the computed CCA value through the periodic transmission of beacon frames  (one beacon frame per 100 msec is transmitted with our ipw2200 driver) \cite{mdg}. 
As one might expect, the jammer's signal may collide with beacon frames, and this makes it more difficult for the power control module to converge. 
Note also that as reported in \cite{mdg, imc05vasud}, beacon transmissions are not always timely, especially in conditions of high load and poor-quality links (such as in jamming scenarios). 
We measure the network-wide convergence time, i.e., the time elapsed from the moment that we activate the jammer until all legitimate devices have adjusted their CCA threshold as per our power control scheme. 
First, we perform measurements on a multi-hop mesh topology consisting of 5 APs and 10 clients (2 clients per AP, all equipped with the {\em Intel 2915} cards). 
In order to have an idea about whether the observed convergence time is significant, we also perform experiments without jammers, wherein we manually invoke the power control module through a user-level socket interface on one of the APs. 
We observe that the convergence time for the specific setting is approximately 1.2 sec. 
Then, we activate a {\em deceptive} jammer in a close proximity to 2 neighbor APs (MRC is disabled; the jammer affects only the 2 APs). 
Table \ref{tab:powConver} contains various average convergence times for the specific setting and for different $P_J$ values. 
\begin{table}[h]
\vspace{-0.05in}
\centering \small
\begin{tabular}{|c|c|}
\hline $P_J$ (dB) & Convergence time (sec) \\ 
\hline 1 & 1.8  \\ 
\hline 2 & 2.4  \\ 
\hline 3 & 2.8  \\ 
\hline 4 & 3.5  \\ 
\hline 
\end{tabular} 
\caption{Average convergence times (in sec) for different $P_J$ values.}
\label{tab:powConver}
\vspace{-0.08in}
\end{table}
\\
We observe that even though the convergence time increases due to jamming, it still remains rather short. 
Furthermore, we perform extensive experiments with 8 APs, 19 clients and 4 balanced jammers with $P_J$ = 3 dBm, all uniformly deployed.  
We observe that in its distributed form the power control module converges in approximately 16 sec in our network-wide experiments. 
Although one may expect different (lower or higher) convergence times with different 
hardware/software and/or mobile jammers, these results show that in a static topology the power control module converges relatively quickly.

\vspace{-0.2cm}
\section{Scope of ARES} 
\label{sec:discussion}
\setcounter{paragraph}{0}

From our evaluations, it is evident that ARES can provide  performance benefits in the presence of jamming, 
even with other wireless technologies, and both in static and dynamically changing  environments. 
In this section, we discuss some design choices and the applicability requirements of ARES. 

\textbf{\em ARES does not require additional complicated hardware or software functionalities:}
The two modules that constitute ARES are relatively easy to implement in the driver/firmware of commodity wireless cards, and do not require any hardware changes. 
The only software modification needed in the firmware involves the CCA tuning functionality. 
Specifically, it should be possible to change the CCA threshold as per the commands sent through a driver-firmware socket interface. 
To facilitate a  distributed WLAN implementation of ARES, the AP 
driver needs to be modified to read the new Beacon template from the Beacons received from neighbor co-channel APs. 
Finally, clients need to apply the power and CCA settings determined by their affiliated AP. 


\textbf{\em On the  effectiveness of MRC:} 
Our analysis provided in section \ref{sec:rate} is an accurate tool that decides between the use of a fixed rate or a rate adaptation strategy.  
However, 
applying the analysis in a real system is quite challenging, for various reasons. 
In particular, as discussed earlier the analysis requires a set of inputs which may not be readily available. 
If the analysis were to be applied in real time, 
ARES would need to observe these values on the fly and invoke the rate module whenever significant, non-temporal changes 
are observed. 
It is also difficult to derive the jammer's distribution accurately and quickly. 
Such requirements make the application of the analysis somewhat infeasible in real-time systems. 
Furthermore, the analysis  
can account for the presence of one jammer only. In scenarios with multiple jammers, 
it cannot decide between fixed or adaptive rate.  
%

In contrast, our more practical scheme MRC does not need any inputs. 
It can operate efficiently even with multiple jammers. 
Note that MRC in its current form takes into account the time\footnote{In its current form, this time is in terms of 
the number of jamming cycles; 
this can be easily modified to use more generic time units.} that has elapsed since the last time that rate control was invoked.  
The policy is to invoke the rate adaptation strategy after periodic intervals. The optimal rate at which  rate adaptation 
should be invoked depends on the temporal variability of the channel.
In particular, to perform this optimally, ARES would need to measure (or estimate) the coherence time $\tau$ of the channel (time
for which the channel remains unchanged \cite{Camp08}) and invoke the rate control algorithm every $\tau$ secs.  
While this is not possible with current 802.11 hardware, it may be possible in the future \cite{Camp08}.
Alternatively ARES could employ a learning strategy as discussed in Section \ref{sec:system}.
Enhancing the rate control module to address these issues is in our future plans.

\textbf{\em ARES with reactive and constant jammers:} For the most part in this work we considered various types of  random jammers. 
With constant jammers, rate adaptation is not expected to provide benefits, 
since the 
continuous jamming interference does not allow the use of high rates. 
%
%
%
Nevertheless, rate control (even as a {\em standalone} module) is expected to provide benefits in the presence of reactive jamming.  
In particular, 
let us consider a link consisting of  
legitimate nodes A and B. 
The reactive jammer J needs to sense the ongoing transmission and quickly transmit its jamming signal. 
If we denote by $t_{flight}$, the flight time of the legitimate packet and with $t_{sense}$ the time needed for J to sense this packet, then the probability of succesful packet corruption\footnote{We assume an optimal reactive jammer, i.e., one which is able to jam at the exact time instance when it senses a legitimate packet (best case scenario for the adversary). In reality, this will not be the case.} 
can be calculated as: $P_{jam}=P(t_{sense}<t_{flight})$.  
Assuming that $t_{sensing}$ is uniformly distributed at the interval $[0,DIFS]$\footnote{This is a reasonable assumption to make since the protocol allows for a DIFS period in order to sense any transmission.} we get:
\begin{equation}
P_{jam}=\int\limits_{0}^{t_{flight}} \frac{1}{DIFS}\, dt=\frac{t_{flight}}{DIFS}=\frac{\# bytes/packet}{rate\cdot DIFS}
\label{eq:reactive}
\end{equation}
From Eq. \ref{eq:reactive} it is clear that through the use of high bit rates and/or reduced packet sizes 
the probability of succesful reactive jamming can be decreased.  
However, there is a tradeoff between successful reception and decreased jamming probability that needs to be examined more carefully.  
Finally, the power control module of ARES, 
can be useful in the presence of both constant (as shown in the previous section) and reactive jamming. 
%
%
%
%
%
%
%
%
%
%

\section{Conclusions} 
\label{sec:conclusions}
\setcounter{paragraph}{0}

We design, implement and evaluate ARES, an anti-jamming system for 802.11 networks. 
ARES has been built based on observations from extensive measurements on an indoor testbed in the presence of random jammers,  
and  is primarily composed of two  modules. 
The {\em power control module} tunes the CCA thresholds  
in order to allow the transmission and capture of legitimate packets in the presence of  the jammer's signals,  to the extent possible. 
The {\em rate control module} decides between fixed or adaptive-rate assignment. 
We demonstrate the effectiveness of ARES in three different deployments 
{\bf (a)} a 802.11n based MIMO WLAN, 
{\bf (b)} a mesh network infested with mobile jammers, and 
{\bf (c)} a 802.11a WLAN with uplink TCP traffic. 
ARES can be used in conjunction with other jamming mitigation techniques (such as frequency hopping or directional antennas).
Overall, the application of 
ARES leads to significant performance benefits in jamming environments. 


\section*{Acknowledgments}
The authors would like to thank Ralink Technologies for providing us the Linux source driver for the RT2860 AP and Intel Research, for providing us with the prototype firmware of {\em ipw2200} AP.

\vspace*{0.5mm}
\small
\bibliographystyle{unsrt}
\bibliography{main}

\begin{thebibliography}{10}

\bibitem{sesp}
{SESP jammers}.
\newblock http://www.sesp.com/.

\bibitem{conf-jam}
{Jamming attack at hacker conference}.
\newblock http://findarticles.com/p/articles/mi\_m0EIN/
  is\_2005\_August\_2/ai\_n14841565.

\bibitem{xbox-jam}
{Techworld news}.
\newblock http://www.techworld.com/mobility/ news/index.cfm?newsid=10941.

\bibitem{rf-jam}
{RF Jamming Attack}.
\newblock http://manageengine.adventnet.com/
  products/wifi-manager/rfjamming-attack.html.

\bibitem{navda07}
V.~Navda, A.~Bohra, S.~Ganguly, and D.~Rubenstein.
\newblock {{Using Channel Hopping to Increase 802.11 Resilience to Jamming
  Attacks}}.
\newblock In {\em IEEE INFOCOM mini-conference}, 2007.

\bibitem{Xu04}
W.~Hu, T.~Wood, W.~Trappe, and Y.~Zhang.
\newblock {{Channel Surfing and Spatial Retreats: Defenses Against Wireless
  Denial of Service}}.
\newblock In {\em ACM Workshop on Wireless Security}, 2004.

\bibitem{wide-jam}
{ISM Wide-band Jammers}.
\newblock http://69.6.206.229/e-commerce-solutions-catalog1.0.4.html.

\bibitem{intech-jam}
{ISA: Users fear wireless networks for control}.
\newblock http://lists.jammed.com/ISN/2007/05/0122.html.

\bibitem{kpele-wiopt09}
K.~Pelechrinis, C.~Koufogiannakis, and S.V. Krisnamurthy.
\newblock {{Gaming the Jammer: Is Frequency Hopping Effective?}}
\newblock In {\em WiOpt}, June 2009.

\bibitem{ieee80211}
IEEE~Std 802.11g Part 11: Wireless LAN Medium Access Control~(MAC) and Physical
  Layer~(PHY) specifications.
\newblock 2003.

\bibitem{powInfocom}
V.~Mhatre, K.~Papagiannaki, and F.~Baccelli.
\newblock {{Interference Mitigation through Power Control in High Density
  802.11 {WLAN}s}}.
\newblock In {\em IEEE INFOCOM}, 2007.

\bibitem{bicket}
J.~Bicket.
\newblock {{Bit-rate Selection in Wireless Networks}}.
\newblock In {\em MS Thesis, Dept. of Electr. Engin. and Comp. Science, MIT},
  2005.

\bibitem{onoe}
{Onoe Rate Control}.
\newblock http://madwifi.org/browser/trunk/ath\_rate/onoe.

\bibitem{amrr}
S.~Pal, S.~R. Kundu, K.~Basu, and S.~K. Das.
\newblock {{IEEE 802.11 Rate Control Algorithms: Experimentation and
  Performance Evaluation in Infrastructure Mode}}.
\newblock In {\em PAM}, 2006.

\bibitem{Xu05}
W.~Xu, W.~Trappe, Y.~Zhang, and T.~Wood.
\newblock {{The Feasibility of Launching and Detecting Jamming Attacks in
  Wireless Networks}}.
\newblock In {\em ACM MOBIHOC}, 2005.

\bibitem{Gummadi07}
R.~Gummadi, D.~Wetheral, B.~Greenstein, and S.~Seshan.
\newblock {{Understanding and Mitigating the Impact of RF Interference on
  802.11 Networks}}.
\newblock In {\em ACM SIGCOMM}, 2007.

\bibitem{Xu06}
W.~Hu, K.~Ma, W.~Trappe, and Y.~Zhang.
\newblock {{Jamming Sensor Networks: Attacks and Defense Strategies}}.
\newblock In {\em IEEE Network}, May/June 2006.

\bibitem{Lin03}
G.~Lin and G.~Noubir.
\newblock {{On Link Layer Denial of Service in Data Wireless LANs}}.
\newblock In {\em Wireless Communications and Mobile Computing}, May 2003.

\bibitem{Noubir03}
G.~Noubir and G.~Lin.
\newblock {{Low-power DoS Attacks in Data Wireless LANs and Countermeasures}}.
\newblock In {\em ACM MOBIHOC (poster)}, 2003.

\bibitem{Noubir}
G.~Noubir.
\newblock {{On Connectivity in Ad Hoc Network under Jamming Using Directional
  Antennas and Mobility}}.
\newblock In {\em Wired/Wireless Internet Communications, Vol. 2957/2004, pp.
  186-200, 2004}.

\bibitem{wong06}
S.Wong, H.Yang, S.Lu, and V.Bharghavan.
\newblock {{Robust Rate Adaptation in 802.11 Wireless Networks}}.
\newblock In {\em MOBICOM}, 2006.

\bibitem{shah05}
V.Shah and S.V.Krisnamurthy.
\newblock {{Handling Assymetry in Power Heterogeneous Ad hoc Networks: A Cross
  Layer Approach}}.
\newblock In {\em IEEE ICDCS}, 2005.

\bibitem{trid}
I.~Broustis, J.~Eriksson, S.~V. Krishnamurthy, and M.~Faloutsos.
\newblock {{A Blueprint for a Manageable and Affordable Wireless Testbed:
  Design, Pitfalls and Lessons Learned}}.
\newblock In {\em IEEE TRIDENTCOM}, 2007.

\bibitem{ucrtestbed}
{UCR Wireless testbed}.
\newblock http://networks.cs.ucr.edu/testbed.

\bibitem{madwifidriver}
{The MadWiFi driver}.
\newblock http://madwifi-project.org/.

\bibitem{ralinkdriver}
{RT2860 wireless driver}.
\newblock http://www.ralinktech.com/ralink/Home /Support/Linux.html.

\bibitem{etx}
D.~S. J.~De Couto, D.~Aguayo, J.~Bicket, and R.~Morris.
\newblock {{A High Throughput Path Metric for MultiHop Wireless Routing}}.
\newblock In {\em ACM MOBICOM}, 2003.

\bibitem{atheros-paper}
J.~C. Chen and J.~M. Gilbert.
\newblock {{Measured Performance of 5-GHz 802.11a Wireless LAN Systems.}}
\newblock In {\em Atheros Comm. White Paper}, August 2001.

\bibitem{mdg}
I.~Broustis, K.~Papagiannaki, S.~V. Krishnamurthy, M.~Faloutsos, and V.~Mhatre.
\newblock {{MDG: Measurement-Driven Guidelines for 802.11 WLAN Design}}.
\newblock In {\em ACM MOBICOM}, 2007.

\bibitem{bug}
{SampleRate Bug}.
\newblock http://madwifi.org/ticket/989.

\bibitem{zvan}
S.~Zvanovec, P.~Pechac, and M.~Klepal.
\newblock {{Wireless LAN Networks Design: Site Syrvey or Propagation Models?}}
\newblock In {\em Radioengineering, Vol. 12, No. 4}, Dec. 2003.

\bibitem{jafar}
H.Jafarkhani.
\newblock {\em {{Space-Time Coding: Theory and Practice}}}.
\newblock Cambridge University Press, 2005.

\bibitem{imc05vasud}
S.~Vasudevan et~al.
\newblock {Facilitating Access Point Selection in IEEE 802.11 Wireless
  Networks}.
\newblock In {\em ACM IMC}, 2005.

\bibitem{Camp08}
J.~Camp and E.~W. Knightly.
\newblock {Modulation Rate Adaptation in Urban and Vehicular Environments:
  Cross-Layer Implementation and Experimental Evaluation}.
\newblock In {\em ACM MOBICOM}, 2008.

\end{thebibliography}

\end{document}